\documentclass[a4paper,11pt]{article}
\pdfoutput=1 

\usepackage{jcappub} 
\usepackage[T1]{fontenc} 
\usepackage{subcaption}
\usepackage{pgfplots}
\usepackage{hyperref}
\usepackage{amsmath}
\usepackage{multirow}
\usepackage{placeins}
\usepackage{gensymb}
\newcommand*{\doi}[1]{\href{http://dx.doi.org/#1}{doi: #1}}
\DeclareMathOperator*{\aggfn}{\square}
\title{\boldmath Application of Graph Networks to background rejection in Imaging Air Cherenkov Telescopes}

\author[a]{J. Glombitza,}
\author[a]{V. Joshi,}
\author[a]{B. Bruno,}
\author[a]{and S. Funk}

\affiliation[a]{Friedrich-Alexander-Universit\"at Erlangen-N\"urnberg, Erlangen Centre for Astroparticle Physics, Nikolaus-Fiebiger-Str. 2, 91058 Erlangen, Germany}

\emailAdd{jonas.glombitza@fau.de}
\emailAdd{vikas.joshi@fau.de}

\abstract{
Imaging Air Cherenkov Telescopes (IACTs) are essential to ground-based observations of gamma rays in the GeV to TeV regime. One particular challenge of ground-based gamma-ray astronomy is an effective rejection of the hadronic background. 
We propose a new deep-learning-based algorithm for classifying images measured using single or multiple Imaging Air Cherenkov Telescopes.
We interpret the detected images as a collection of triggered sensors that can be represented by graphs and analyzed by graph convolutional networks.
For images cleaned of the light from the night sky, this allows for an efficient algorithm design that bypasses the challenge of sparse images in deep learning approaches based on computer vision techniques such as convolutional neural networks. We investigate different graph network architectures and find a promising performance with improvements to previous machine-learning and deep-learning-based methods.}

\arxivnumber{2305.08674} 

\begin{document}
\maketitle
\flushbottom

\section{Introduction}
\label{sec:intro}
Arrays of Imaging Air Cherenkov Telescopes (IACTs) enable precise observations of the very-high-energy gamma-ray sky in the GeV to the TeV regime.
When highly-energetic particles penetrate the Earth's atmosphere, air showers (i.e., particle cascades) are initiated. IACTs image the Cherenkov radiation, which the relativistic secondary particles emit while passing through the Earth's atmosphere.

For the precise observation of gamma-ray emitters, the hadronic background has to be reduced to a minimum to ensure a high sensitivity of the instruments. For rejecting this background --- of mainly proton-induced air showers --- most algorithms rely on high-level image observables like Hillas parameters~\cite{hillas}.
In this approach, the images are first cleaned from background noise, e.g., the night sky background.
Subsequently, the images --- consisting of the remaining pixels after cleaning --- are reduced to multiple parameters by elliptical modeling of the first moments of the Cherenkov light distribution, where the width and length of the ellipse are critical parameters and decisive for $\gamma$/hadron separation.
Since hadron-induced showers are subject to larger fluctuations due to the hadronic interactions in the shower development, the resulting light distribution is different and mainly results in increased widths.
For an efficient background rejection, cut strategies use these differences on such image parameters or by exploiting their correlations using machine learning techniques~\cite{OHM2009383}.

With the recent advent of deep learning~\cite{LeCun2014, Goodfellow-et-al-2016}, the application of machine-learning algorithms to low-level data of physics instruments became computationally feasible~\cite{dlfpr}. These deep learning techniques already demonstrated their effectiveness in Large Hadron Collider (LHC)~\cite{Guest_2018}, neutrino~\cite{Aurisano_2016, Abbasi_2021}, and cosmic-ray physics~\cite{Aab:2021rcn, Aab_2021}.
Also, in gamma-ray astronomy, various works focus on reconstructing IACT images using deep learning, i.e., neural networks, to extract information beyond the Hillas parameterization, which is too simplified to describe characteristics of the Cherenkov light distribution in full detail.
For example, especially proton showers that develop on average far more inhomogeneously compared to photon showers cannot properly be described by an ellipse. A simple analysis based on width and length can thus be interpreted as a principal component analysis (PCA) where only the first two components of the light distribution are considered.
Hence, more sophisticated approaches aim to exploit the substructure of IACT images beyond these moments to improve the rejection of hadron-initiated showers.

\begin{figure}[t!]
    \begin{subfigure}[b]{0.5\textwidth}
        \centering
        \includegraphics[height=0.9\textwidth]{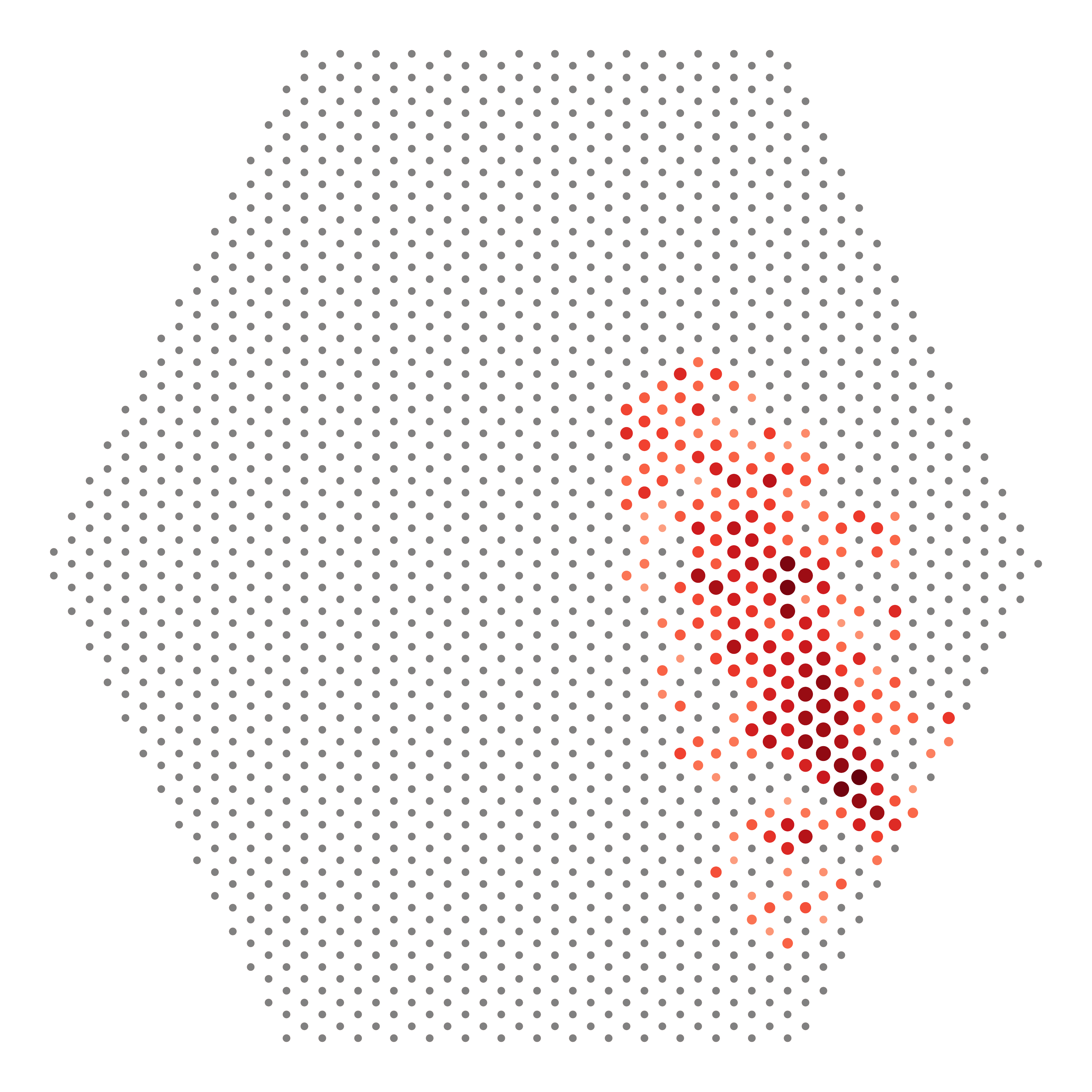}
        \caption{}
        \label{fig:raw_event}
    \end{subfigure}%
    \begin{subfigure}[b]{0.5\textwidth}
        \centering
        \includegraphics[height=0.9\textwidth]{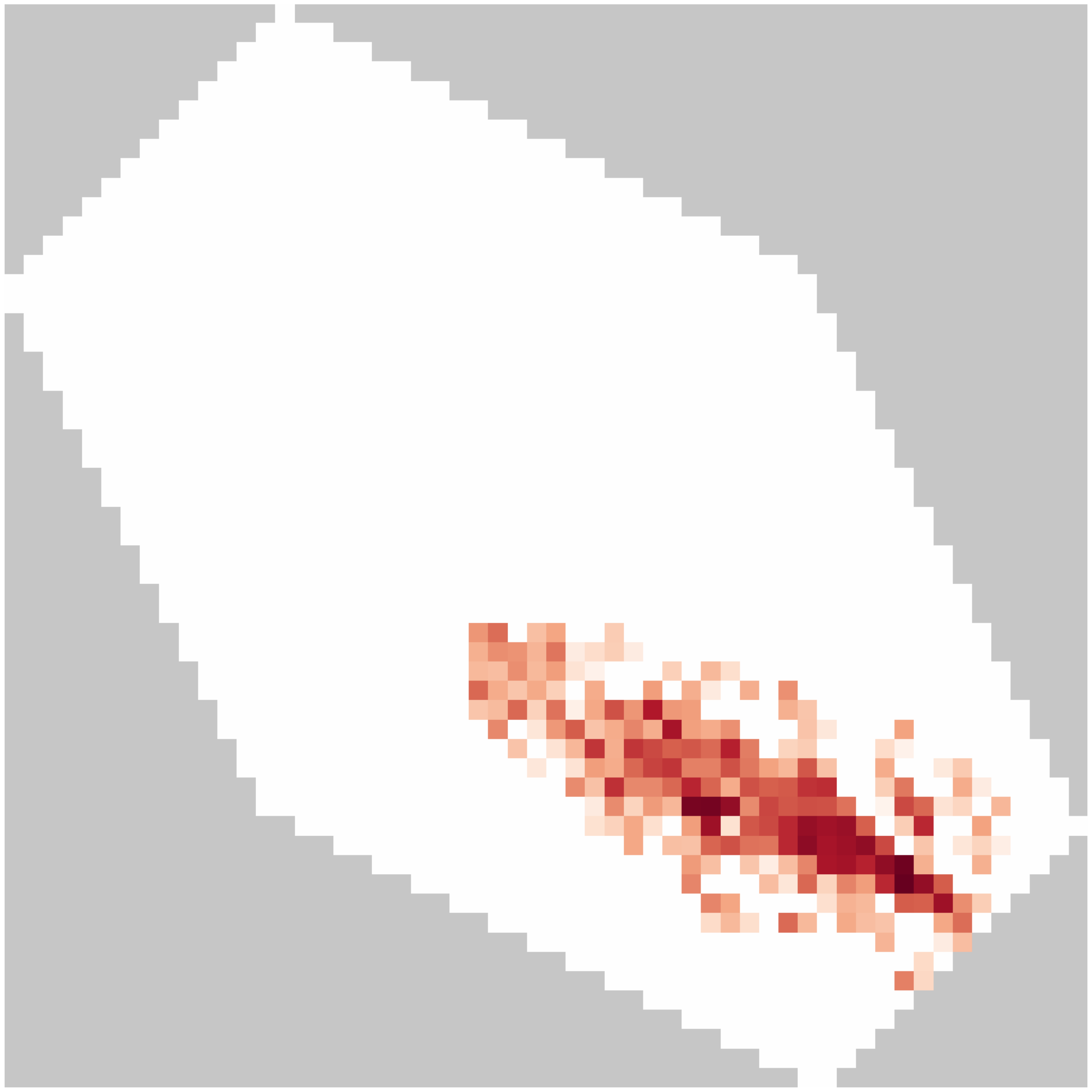}
        \caption{}
        \label{fig:as_image}
    \end{subfigure}
\caption{Representation of a typical event measured using CT5. (a) Detected Cherenkov light distribution at the camera after cleaning and pre-processing (point cloud). Marker sizes and colors indicate the signal strength. Grey points represent pixels without signal after cleaning and pre-processing. (b) Representation as a Cartesian image using axial addressing; grey pixels mark zero-padded regions, and white pixels indicate pixels without any signal after cleaning and pre-processing.}
\label{fig:img}
\end{figure}

The first application of deep learning to IACT images was studied in Ref.~\cite{Feng_2016}. A following comprehensive study on the application of deep learning to event reconstruction~\cite{Shilon_2019} forms the base for many IACT deep learning applications~\cite{ct_learn, Brill_2019, Jacquemont_2021, Spencer_2021}.
The proposed hybrid architecture consisting of convolutional and recurrent operations showed promising results when applied to simulations.
Whereas the recurrent network part was used to deal with image stereoscopy, i.e., the combination of multiple telescopes that imaged the same shower, the convolutional network part was used to exploit the shower image on a per-telescope basis.
One particular challenge of this approach is the image sparsity, as after image cleaning, typically only around $10-15\%$ active pixels remain, as visible in Figure~\ref{fig:raw_event}.
Since many IACT cameras feature a hexagonal arrangement of pixels, the sparsity of the image even increases as a transformation into a Cartesian representation is needed (compare Figure~\ref{fig:as_image}). These caveats make the application of convolutional neural networks (CNNs) less efficient and motivate algorithms that can deal more naturally with the measured light distribution in the camera.
When not performing cleaning, in theory, more information could be exploited assuming that the cleaning has finite performance. Nevertheless, the major part of the image would, on average, consist of background noise that must be described in full detail. We, therefore, would like to emphasize that investigating additional more sophisticated cleaning procedures will also contribute decisively to the development of precise and efficient algorithms for IACT image analyses.

In this work, we present a novel approach for discriminating $\gamma$-ray- from proton-induced showers using graph networks, which were recently applied to physics data lying on non-regular~\cite{icecube_class, TAGConvIceCube, Abbasi_2022} and non-Euclidean manifolds~\cite{Qu_2020, Bister_2021}.
Our technique relies on cleaned images, which we interpret as sparse signal patterns that can be represented by graphs, and, thus, efficiently analyzed using graph convolutional neural networks. Furthermore, since only the signal pixels after cleaning contribute to the construction of graphs, our algorithm is computationally significantly more efficient in terms of memory consumption and run time than current approaches based on CNNs or CNN-RNN-based models.

The work is structured as follows. In the first part, we introduce the data, followed by a summary of deep-learning-based IACT image analyses and our new approach to interpreting IACT images as graphs.
Then, in Section~\ref{sec:gnn}, we review two different graph convolutional neural network architectures we use in this work and describe the training of our algorithm using simulations of the High Energy Stereoscopic System (H.E.S.S.)~\cite{hess_crab}.
Finally, in Section~\ref{sec:analysis}, we apply our method to the classification of mono and stereo events, examine the performance of our proposed classification algorithm, and compare the results to machine-learning and deep-learning-based classifiers presented in the literature.

\section{Data}
\label{sec:data}
For the simulation of extensive air showers, the software package CORSIKA (Cosmic Ray
Simulation for Kascade) is used~\cite{heck_1998}. Initially developed for the KASCADE experiment, it is publicly available, open-source, and a standard tool for the broader astroparticle physics community.
Furthermore, we use the software \texttt{sim\_telarray} to fully simulate the detector response, ranging from the photons’ ray tracing to the measurement with photomultiplier tubes (PMTs) and its digitization~\cite{Bernl_hr_2008}. 

The simulated events are calibrated and cleaned using the standard H.E.S.S. Analysis Program (HAP). H.E.S.S. is one of the currently operational IACT arrays and is located in Namibia. It consists of five telescopes, four small telescopes, named CT1-4, arranged in a square with 120~m side length, and a larger fifth telescope (CT5) placed in its center. In this work, we use H.E.S.S. as a show-case scenario of our algorithm. Nevertheless, due to its flexible design, our proposed method can be applied to any IACT array configuration.

Once the data is recorded, a calibration procedure is needed to use the raw data, which entails converting the measured Analog-to-Digital-Converter (ADC) counts into the physical unit of photoelectrons (p.e.). In the following step, a dual threshold cleaning procedure is applied to exclude all pixels without a shower signal and to keep only pixels with shower signals in local proximity. 
We used the so-called extended 4/7 cleaning~\cite{hess_crab} for CT1-5 in this paper. This means, during the tail cuts cleaning process, we keep a pixel if the intensity is >4~p.e. and at least one nearest neighbor pixel has an intensity >7~p.e., conversely. Finally, we save the extended image, i.e., also include the neighboring four rows of pixels around the 4/7-cleaned image.
We further use selection cuts, part of the standard analysis chain of H.E.S.S., selecting only bright camera images that are not truncated, ensuring high-quality reconstructions. Whereas the minimum image total amplitude, denoted as the image size parameter, guarantees bright images, the local distance, i.e., the maximum distance between the Hillas ellipse center-of-gravity and the camera center, ensures that the images are not truncated. The local distance cut, we refer to it as \emph{preselection} cut hereafter.

In H.E.S.S., the reconstruction is performed with different configurations of the five telescopes enabling observations at different energy ranges. These configurations are motivated by either the start of operation of individual telescopes -- for example -- CT1-4 have been operational since 2003, and CT5 was added to the array in 2012. At the time of writing, there are three main configurations: \emph{mono},  \emph{stereo}, and \emph{hybrid}. "Mono", corresponds to observations that solely include CT5 -- most effective at low energies. The "stereo" and "hybrid" configuration corresponds to any combination of at least two telescopes from CT1-4 and CT1-5, respectively, having good quality data, i.e., satisfying the required selection. Whereas stereo is most effective at the highest energies, hybrid covers the entire energy range of H.E.S.S.

In this work, for training the neural networks, all events were simulated as diffuse emission around $20\degree$ zenith, $0\degree$ azimuth, with an opening angle of $5\degree$. The simulations were performed for the so-called \texttt{phase2d3}, which corresponds to the latest state of the H.E.S.S. array taking into account effects such as the optical efficiency of the telescope mirrors, in an energy range of 10~GeV to 300~TeV with a spectral index of -2. The proton showers were simulated using the \texttt{QGSJET-II} model implemented in CORSIKA.
Since the amount of Cherenkov light produced differs between photon and proton showers due to the hadronic component, the simulated energy range differs slightly between both classes to cover the full sensitivity range of H.E.S.S.\footnote{For photon showers, the range is 10~GeV to 100~TeV, and for protons 10~GeV to 300~TeV. We further checked that the distributions of image sizes and reconstructed energies of proton and photon showers are similar. This important step ensures that the trained classifiers are unbiased, i.e., are sensitive to differences in the image light distribution and not the energies.}.
We have pre-processed the simulated events using HAP to apply the image cleaning, trigger condition for the array, and calculate the Hillas parameters per image in mono and hybrid analysis in the so-called \texttt{zeta} configuration~\cite{OHM2009383}.

For detailed comparisons, we have also used a standard set of cuts, where the minimum size parameter of an image is $60 \, \mathrm{p.e.}$ for CT1-4 and $80 \, \mathrm{p.e.}$ for CT5 and the maximum local distance of an image is $0.525 \, \mathrm{m}$ for CT1-4. For CT5, we do not apply a local distance cut.
Our final data consists of $10^6$ events in the mono data set, $650{,}000$ events in the stereo data set with preselection, and $10^6$ events for the stereo dataset without the preselection, i.e., the local distance cut applied.
We used $85\%$ for validation and training and $15\%$ for final testing.

Finally, we compare  our results as a benchmark to the BDT-based standard $\gamma$/hadron separation method used in H.E.S.S. However, we note that the BDT-based method used in H.E.S.S. requires full preselection and further uses point-source gamma-ray simulations, and real "off-run" data is used for the background description. These off-runs contain the data collected by pointing to the parts of the sky where we do not have any known gamma-ray source, enabling a stable application of the BDT to data. For our scenario, this is a conservative benchmark as we likely overestimate the BDT performance on simulations since, for evaluation of graph networks, diffuse $\gamma$'s were used. 


\subsection{IACT images and graphs}
Current deep-learning approaches for IACT-image analyses are based on CNNs.
This approach --- even if well-motivated and showing promising results --- suffers from the following two challenges.
Firstly, the sparsity gets even more prominent for cleaned images because image cleaning removes a significant fraction of the night-sky background. We found that, on average, only $\sim 10 - 15\%$ of pixels hold signals. See Figure~\ref{fig:raw_event} for a visualization of a typical IACT image measured using CT5 after cleaning and pre-processing.
Secondly, the design of IACT cameras usually features an arrangement of hexagonal pixels. Thus, re-indexing~\cite{ERDMANN201846, hexaconv} or re-binning~\cite{Shilon_2019} has to be performed to allow for using CNNs, which are based on Cartesian grids. It can lead to performance issues for too low-resolution re-binning or inefficient, i.e., sparse representations~\cite{nieto_hex}. For example, the commonly used axial representation introduces a lot of zeros into the new images indicated as grey pixels in Figure~\ref{fig:as_image}.\footnote{The offset representation does not have this inefficient representation. However, it is limited to kernels applied with a stride of 2 to conserve translational invariance.} For our used data, the sparsity would be quite high in a CNN and only around $7\%$ of pixels would feature non-zero signals.
Therefore, to reduce the computational complexity, we follow a new, more natural approach.

We consider the cleaned IACT images as point clouds, i.e., a collection of triggered pixels with positions $(x,y)$ in the camera frame, yielding a signal value $s$ in units of p.e.
To conduct an efficient deep learning approach based on convolutional operations\footnote{Similar as used in CNNs.}, we use the point cloud to create a graph --- i.e., a collection of nodes that are connected via edges --- to be analyzed by graph convolutional networks.

\begin{figure}[t!]
    \begin{subfigure}[b]{0.5\textwidth}
        \centering
        \includegraphics[width=0.99\textwidth]{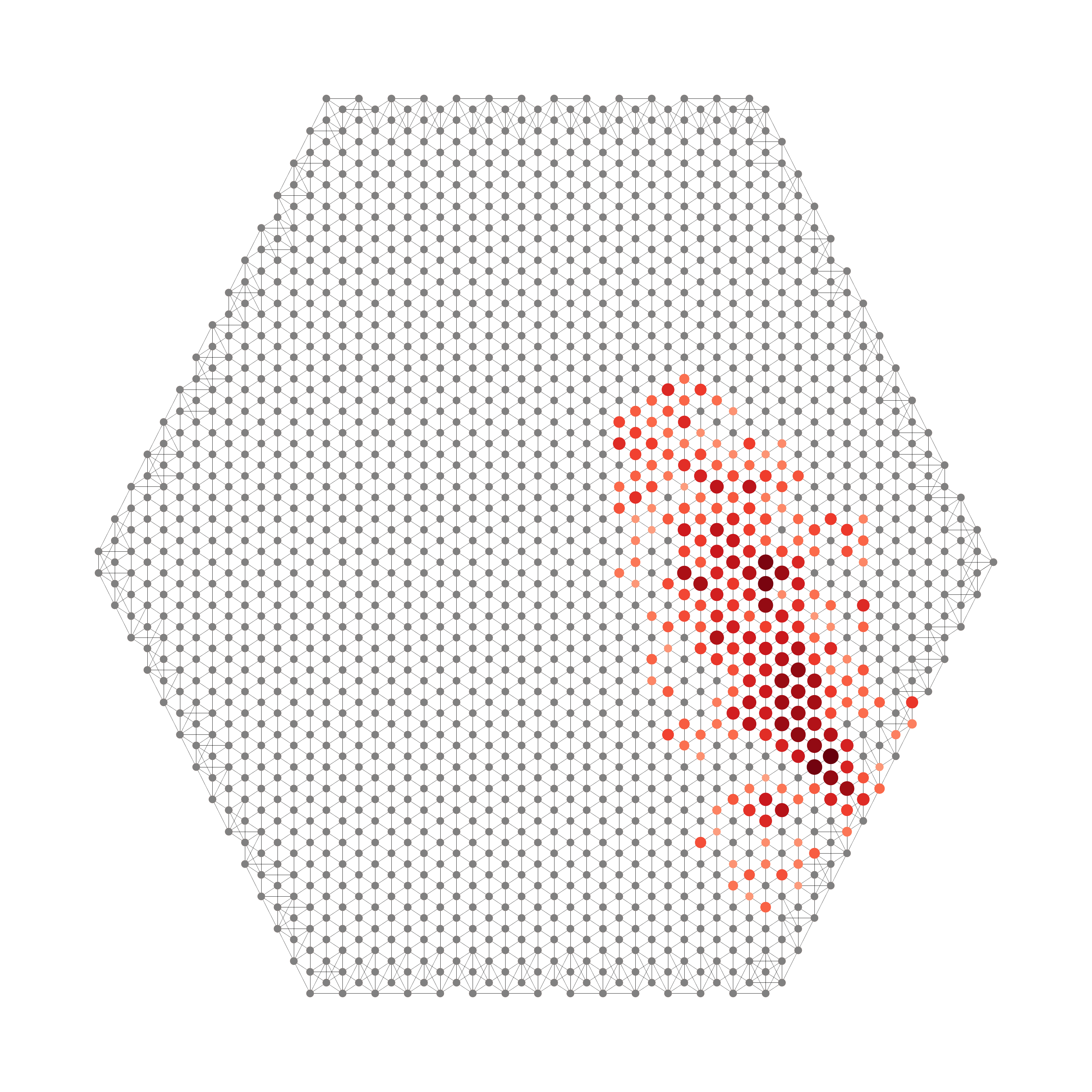}
        \caption{}
        \label{fig:as_fixed_graph}
    \end{subfigure}%
    \begin{subfigure}[b]{0.5\textwidth}
        \centering
        \includegraphics[width=0.99\textwidth]{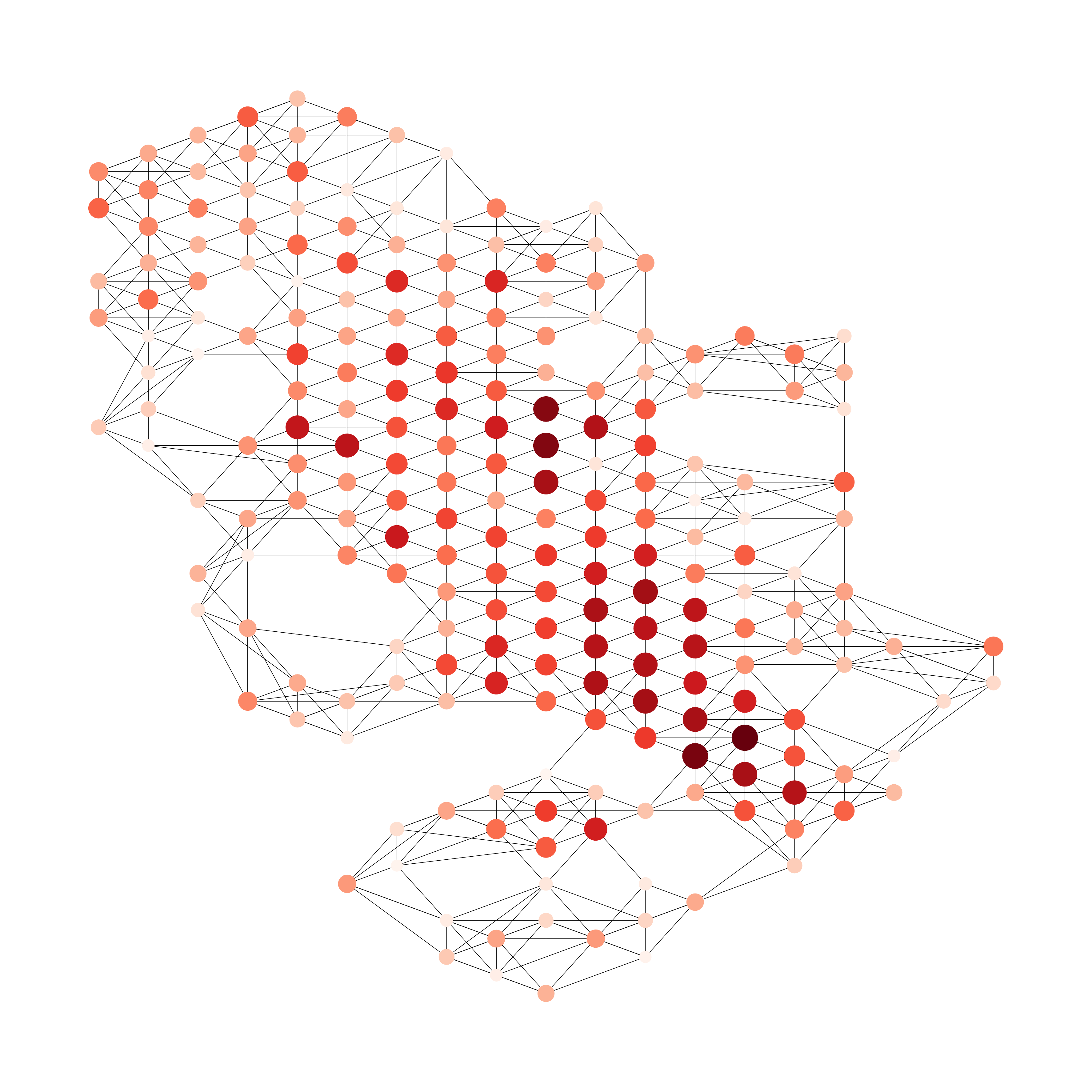}
        \caption{}
        \label{fig:graph}
    \end{subfigure}
\caption{Representation of an IACT image as (a) fixed graph and (b) signal graph. Node sizes denote signal strength. Grey nodes indicate pixels without any signal after pre-processing. The black edges denote the neighborhood relations after constructing the graph using $k$NN. For better visibility, self-loops are not visualized.}
\label{fig:graph_reps}
\end{figure}

As IACT images can be interpreted as an overlay of images from different stages of shower development, the spatial structure of these recorded images in the camera frame is the fundamental feature of discriminating protons from photons. 
Thus, we construct a directed graph using the $k$-nearest-neighbor ($k$NN) algorithm applied to the pixel positions $(x,y)$ only. Due to the hexagonal pixelization, we use $k=6$ and include a self-loop so that each sensor is connected to its closest six neighbors and itself. In principle, two approaches are possible, using all pixels and building a single fixed graph for all events or by only considering the signal pixels to form a signal graph, which changes on an event-by-event basis, as shown in Figure \ref{fig:graph_reps}. We are focusing in this work on the signal graph approach due to its computational efficiency. We also performed a few tests using the fixed graph approach but observed non-significant improvements.

The final image representation results in an un-weighted signal graph with $N$ nodes (remaining number of pixels after cleaning), depicted in Figure~\ref{fig:graph}. Each node $\mathbf{x}_i$ of the graph nodes $\{\mathbf{x}_1, . . . , \mathbf{x}_n\}$ holds a three-dimensional feature vector with the $x,y$ position and measured signal $\mathbf{x}_i = (x_i,y_i,s_{i})$ forming the matrix of node feature vectors $\mathbf{X} \in \mathbb{R}^{N \times 3}$.
The edges of the graph indicate a direct neighborhood between two nodes ($0$ or $1$). This relationship is described by the adjacency matrix $\mathbf{A} \in \mathbb{R}^{N \times N}$. The degree matrix $\mathbf{D} \in \mathbb{R}^{N \times N}$ denotes the number of neighbors for each node.

Note that for a stereoscopic event of $j$ triggered telescopes, $j$ independent graphs are created. In contrast, non-triggered telescopes are modeled as a single-node graph with zero elements as the feature vector and a self-loop. This approach simultaneously facilitates an efficient representation and the information on the telescopes' trigger state; the latter is vital for inferring shower properties.

\paragraph{Transformation of signals}
\label{sec:transform}
The distribution of measured pixel signals features an exponential tail towards high values. Therefore, to improve the learning behavior of neural networks, we use a logarithmic re-scaling~\cite{Aab:2021rcn} of the measured pixel signals $s_i$:
\begin{align}
s_i' &= \log_{10}(s_i + 1) \\
s_{i, \mathrm{norm}} &= \frac{s'_i}{\sigma(s')},
\end{align}
where $\sigma(s')$ denotes the standard deviation of the transformed signals estimated over the entire dataset.
Pixels with negative values are removed, as they did not show a performance improvement and reduced the computational efficiency of the algorithm.

\section{Graph Convolutional Neural Networks}
\label{sec:gnn}
In recent years, the success of deep learning in speech recognition and computer vision has spread into the natural sciences, including physics.
One of the driving forces of deep learning is convolutional neural networks (CNNs), which have been demonstrated to be enormously powerful when applied to regular and structured data such as images and time series.
For the analysis of data lying on irregular grids or non-Euclidean manifolds and forming point clouds, however, the convolutional operations have to be substantially modified. Graph convolutional neural networks offer an elegant way out of this challenge.

By constructing graphs to represent these point clouds, a clear neighborhood relation is defined, enabling the exploitation of the data structure using convolutions. Currently, two different approaches are used to perform the convolution operation on graphs. One ansatz, the so-called spectral convolutions, utilizes a transformation into the Fourier domain where the convolution acts point-wise. A second is a spatial approach in which the convolutional kernels are spatially localized --- similar to CNNs --- and the kernels are extended to non-Euclidean and irregular lattices. For a detailed introduction to deep learning on graphs and the different methods, refer to Ref.~\cite{geo_dl, dlfpr}. 

In this work, we investigate two different formulations discussed in Sec.~\ref{sec:edgeconv} and Sec.~\ref{sec:tagconv} that use the spatial ansatz, which is very flexible and has already been successfully utilized in physics~\cite{Qu_2020, Bister_2021, Benato_2022, Abbasi_2022}.

\subsubsection{Computational considerations and comparison to CNNs}
The computational considerations of a DNN algorithm processing large amounts of data need to regard both the usage of RAM and VRAM (on the GPU). This work, in particular, aims to reduce VRAM usage to make DNN training more efficient.
The largest memory consumption of the graph convolutional network (GCN) is acquired by the adjacency matrix, which consists of $N\times2\times (k + 1)$ non-zero entries, stored in a sparse matrix. As for each of the $N$ nodes, $k$ neighbors have to be saved, as well as a self-loop for each node.
For the CNN, $H \times W$ entries need to be stored, where $H$ and $W$ denote the height and width of the image.
In our case, the RAM usage of the GCN and the CNN is similar for the case of mono observations since $N \approx \frac{7}{100} \cdot {H \times W}$.

The advantage of the GCN approach becomes evident for stereoscopic observations and a large number of channels.
Firstly, in the used H.E.S.S. stereo data, on average, only three out of the four telescopes are triggered. Since for a non-triggered telescope, only a single node with a self-loop is used, the computational cost is negligible, and the memory consumption is reduced compared to a CNN of a similar structure\footnote{An alternative of giving only the data of the triggered telescopes to a DNN and using a flexible operation to combine the processed images, e.g., using an RNN~\cite{Shilon_2019}, is also possible. However, the fact that the telescope is not triggered contains additional information that would not be considered, and an RNN is computationally elaborate to train and is therefore disfavored.}, where an image full of zeros would have to be input.
Thus, the memory consumption is decreased by around $25\%$. For large IACT arrays such as CTA, the decrease in consumption would be significantly larger and around an order of magnitude or larger.

Secondly, for each channel in the GCN approach, $N$ values have to be saved, as the adjacency has to be stored only once. Thus, for $F$ channels, $N \times F$ pixels need to be stored compared to $H \times W \times F$ for a CNN approach.  Future studies considering additional features like pixel-wise time information, e.g., the peak time or signal traces, would, therefore, result in a significant advantage and make such studies computationally feasible with DNNs.

For the allocated VRAM on the GPU, the above-mentioned difference in the scaling of the number of channels (filters) makes the significant difference, as in each convolutional layer, only $N \times F$ entries have to be stored in the GCN. Since usually $F$ is a large number and multiple layers are used, the memory consumption is decreased significantly and roughly reduced to the order of the sparsity. In our case, with a sparsity of $7\%$, the reduction amounts to more than a magnitude, enabling fast and computationally efficient training. In addition, the handling of non-triggered telescopes reduces the memory in a similar fashion as in the case of RAM, making the algorithm, in particular, efficient for CTA.

\subsection{\label{sec:edgeconv}EdgeConv}
For the analysis of point clouds, by means of graph networks, edge convolutions were proposed in Ref.~\cite{wang2019dynamic}.
Given a graph with $N$ nodes $\{\mathbf{x}_1, ...., \mathbf{x}_n\}$ as input holding various features, for each node  $\mathbf{x}_i$, similarly, the convolutional operation

\begin{equation}
    \mathbf{x}^{\prime}_i = \aggfn_{j \in \mathcal{N}(i)} h_{\mathbf{\Theta}}(\mathbf{x}_i \, \Vert \, \mathbf{x}_j - \mathbf{x}_i)
\end{equation}
is performed --- followed by an activation function --- to extract and accumulate local features. Here $h_\mathbf{\Theta}$ denotes an arbitrary function with trainable parameters $\mathbf{\Theta}$, usually parameterized as a feed-forward network. It can be interpreted as a continuous kernel function as it depends on the feature vector of the central node $(\mathbf{x}_i)$ and the local relative feature difference to the neighboring node $(\mathbf{x}_j - \mathbf{x}_i)$.
After performing the operation with each neighboring node $\mathbf{x}_j$, the aggregation operation $\aggfn_{j \in \mathcal{N}(i)}$ over the neighborhood of the node $\mathbf{x}_i$ is performed. Here, the aggregation operation "$\aggfn$" is to be defined by the user. We will use $\frac{1}{6} \sum_{j \in \mathcal{N}(i)}$ throughout this manuscript.

Figuratively speaking, EdgeConv can be seen as an extension of the standard CNN operating on an image to continuously-distributed data. In the CNN, the same kernel slides over the image and is, at each position, `convolved' with the local image patch. That is the point-wise product of the kernel and the pixel and its local neighborhood --- at the current kernel position --- is performed. Hereafter, the results are aggregated to a single value by summing up, which leads to the fact that the structure of the image is preserved.
Similarly, in the edge convolution at each node, the convolution is performed using the same kernel function $h_{\mathbf{\Theta}}$, replacing the discrete filter as used in CNNs with a feed-forward network. Hence, applying the kernel function to each neighboring node of the current node resembles the point-wise kernel-patch product in the CNN. Finally, the information is aggregated using $\aggfn_{j \in \mathcal{N}(i)}$. Within this interpretation, the number of features (outputs) of the kernel network would correspond to the number of kernels used in a CNN layer.

\begin{figure}[ht!]
\centering
\includegraphics[width=0.999\linewidth]{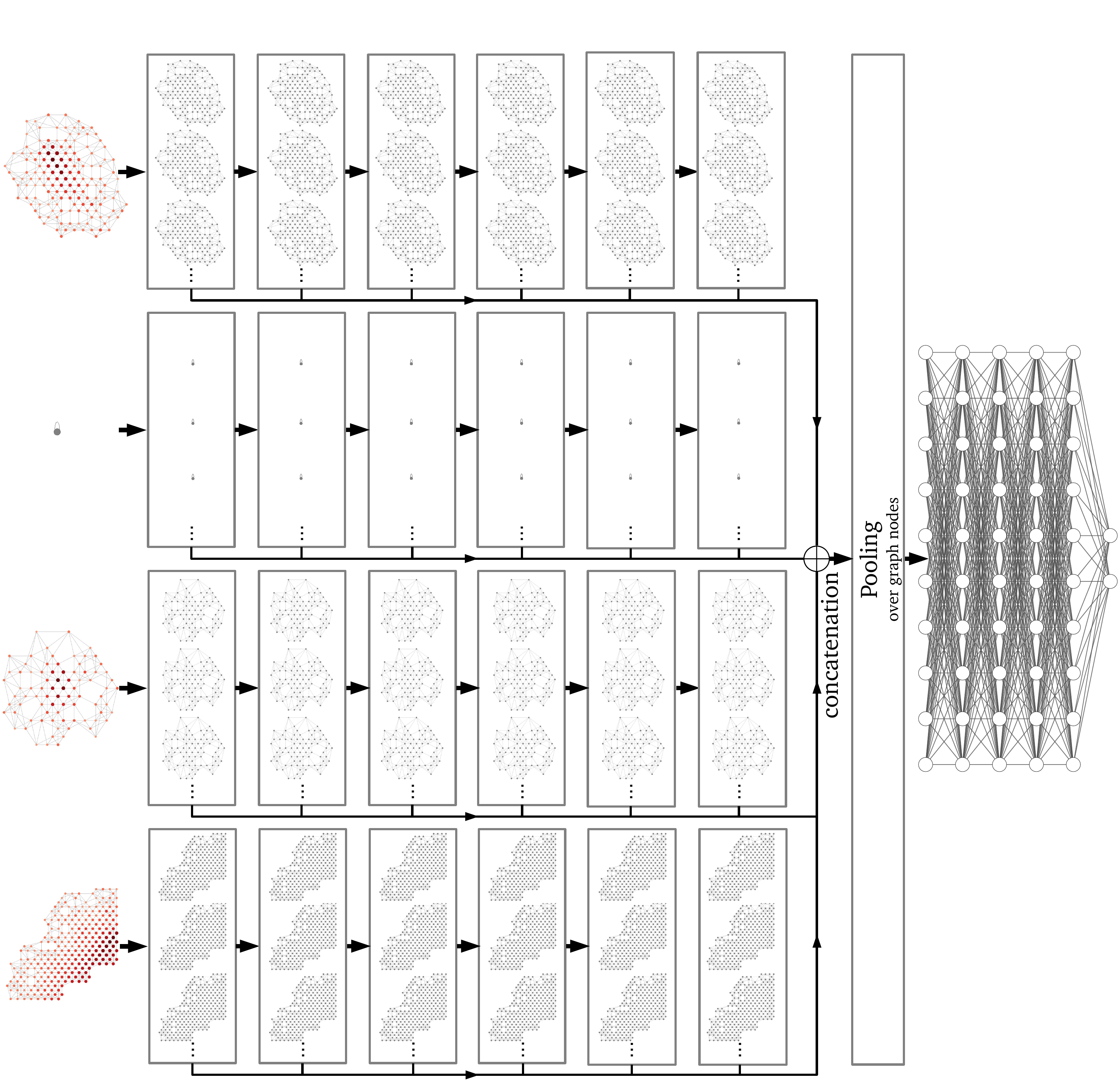}
\caption{Simplified sketch of the graph network design. Here, the more complex version of the stereo architecture is shown. In the visualized event, only three of the four telescopes satisfied the image amplitude cut. After processing the signal patterns of the telescopes using six graph convolutional layers, the resulting features are concatenated, and pooling is performed. Finally, a fully-connected network part follows, analyzing the obtained `pooled' outputs after each convolutional layer of each telescope.}
\label{fig:gnn_architecture}
\end{figure}

\subsection{\label{sec:tagconv}TAGConv}
Topology Adaptive Graph Convolution Networks (TAGCN) were introduced in Ref.~\cite{TAGConv}.
TAGCN extends the well-known fixed-size CNN-like filters to graph-convolutional networks in the spatial domain. In this approach, the graph-convolutional operation is achieved by simultaneously applying a set of fixed-size kernels, to be defined by the user, to each node of a given graph. The output at a given node is each kernel's weighted sum of outputs. The sizes of these kernels are motivated to propagate the information to a given node by its neighbors. For example, size 1 and 2 kernels will bear the information from the next- and next-to-next neighbors, respectively. Note in this setup the number of nodes considered in the convolution depends on the actual size of the one-hop or rather two-hop neighborhood. Thus, it makes the topology of these learnable fixed-size filters adaptive to the topology of the graphs. For a graph with a set of nodes $\mathbf{X}$, the adjacency matrix $\mathbf{A}$, and its degree matrix $\mathbf{D}$, the TAGCN convolution operation is defined as:
\begin{equation}
    \mathbf{X}^{\prime} = \sum_{k=0}^K \left(\mathbf{D}^{-1/2} \mathbf{A} \mathbf{D}^{-1/2}\right)^{k}\mathbf{X} \mathbf{\Theta}_{k},
\end{equation} 
where the summation is taken over the kernel parameter $k$, and $\mathbf{\Theta}_{k}$ are the trainable parameters. More concretely, a given node over which the kernel is applied will only collect the message from the nodes which are $k$ path lengths away, where one path length corresponds to an edge connection between two nodes. In this work, we have used $K=2$, which has also performed best in the original work. 
It is worth emphasizing that the term $\left(\mathbf{D}^{-1/2} \mathbf{A} \mathbf{D}^{-1/2}\right)$ of the adjacency matrix $\mathbf{A}$ is raised to the power $k$, which allows lower contributions in the message passed from a farther node than a nearer one.

For an analogy to CNNs-based architectures where typically a fixed-size filter is used. For example, a $3\times3$ filter in VGG-16~\cite{VGG16} or an $11\times11$ filter in AlexNet~\cite{imagenet}. However, in GoogLeNet~\cite{googlenet}, a combination of a set of filters of different sizes is used in each convolutional layer. Taking a similar approach as in GoogLeNet, TAGCN can be considered a graph-convolution operation where a set of up to $K$-localized filters is used simultaneously. Although we note that in GoogLeNet, the features from different size filters are only concatenated, in contrast, in TAGCN, the weighted sum is aggregated. 

Even being similar, TAGConv is not a simple extension of the well-known graph network architecture~\cite{kipf2017semisupervised}, as it is not an approximation of a graph convolution in the Fourier domain. It can rather be understood in the context of graph signal processing as it utilizes a multiplication by polynomials of the adjacency matrix to consider the $k$-hop neighborhood.

\subsection{Network design and training}
We use a similar layout to enable a fair comparison between the two graph architectures. Each graph part features six graph-convolutional layers of the respective type. For the stereo data set, the same graph part is shared using weight sharing\footnote{The same weights are applied to the signal graphs of the different telescopes.} along the four telescopes as four towers to improve the generalization performance of the architecture.
The following network part receives as input the output of each of the six graph convolutional layers. For the stereo architecture, the outputs of all telescopes are concatenated. To these concatenated tensors, a global pooling operation is applied to remove the dependence on the number of graph nodes. As this is a necessary but aggressive pooling operation, the inputs after each graph layer are utilized to keep information from different levels of the feature hierarchy. The pooling operation is followed by five ResNet modules~\cite{he2015deep} and the output layer.
Figure~\ref{fig:gnn_architecture} shows a sketch of the network design used for the stereo architecture with four towers of graph convolutions processing the signal patterns of the four telescopes. For the mono dataset, the architecture simplifies to a design with only a single tower of graph convolutional layers to process the graph input.

The two graph network architectures were implemented using \texttt{PyTorch Geometric}~\cite{pyg}. For more details on the neural network architectures, refer to section~\ref{sec:gnn_details} in the appendix.
Each network was trained using the ADAM~\cite{kingma_adam_2017} optimizer with the AMSGRAD option~\cite{amsgrad} on a single Nvidia A100 GPU. The training duration amounted to roughly 10-20 hours. The details of the training parameters can be found in the appendix.

Note that for the mono and the stereo data set, we only train on the dataset without the preselection cut applied. We follow this procedure since using the data without preselection increases the statistics by $50\%$. This significant increase in training data will help to improve the generalization performance of our deep learning algorithms.

\section{Classification of IACT images}
\label{sec:analysis}
A crucial aspect of ground-based observation of $\gamma$-rays is to separate the scarce $\gamma$-ray signal events from the abundant hadronic cosmic-ray background events. IACT arrays such as H.E.S.S yield highly-sensitive $\gamma$-ray observations only when the background events are effectively rejected.
Currently, in H.E.S.S, the $\gamma$/hadron separation is performed using so-called Boosted Decision Trees (BDTs). Therefore, we use the BDT classification performance as a baseline comparison for our deep-learning-based classifiers.

In this approach, several parameters extracted from the observed shower images, typically derived from Hillas-based event reconstruction, are combined into one parameter, which describes the likeness of an event to be of hadronic or electromagnetic origin and allows the classification of the events \cite{OHM2009383}. In contrast, our approach directly utilizes the individual pixel-level information mitigating the information loss due to image parameterizations.

In this work, we study the performance of the graph network approach and compare it to the traditional BDT method for two cases: "mono" and "stereo", as described in Section~\ref{sec:data}. 
The networks are evaluated in the respective sensitivity regime of the configuration to obtain realistic results, which are 50~GeV - 300~TeV for mono and 100~GeV - 300~TeV for stereo.

We use the "Receiver Operating Characteristic" (ROC) curve to evaluate the classifiers to demonstrate the general performance. The area under the ROC curve (AUROC) is used as the evaluation metric while comparing different classifiers; as the closer the value of AUROC is to 1, the better the performance of the classifier.

It is to be noted that the $\gamma$/hadron separation is inherently dependent on the amount of information in the recorded images itself. Meaning with an increasing amount of light in the recorded images, the performance of the classification tasks is expected to improve. Thus, we examine the AUROC as a function of the reconstructed energy (derived from the image-amplitude parameter), which is highly correlated with the amount of light observed. To derive the uncertainty on the AUROC value for a given bin in reconstructed energy, we used bootstrapping and resample 1000 times. We also compare the performance to the previous work of deep CNN-RNN-based analysis for the "stereo" configuration~\cite{Shilon_2019}. 

\subsection{Mono performance}
\begin{figure}[t!]
    \begin{subfigure}[b]{0.5\textwidth}
        \centering
        \includegraphics[width=0.99\textwidth]{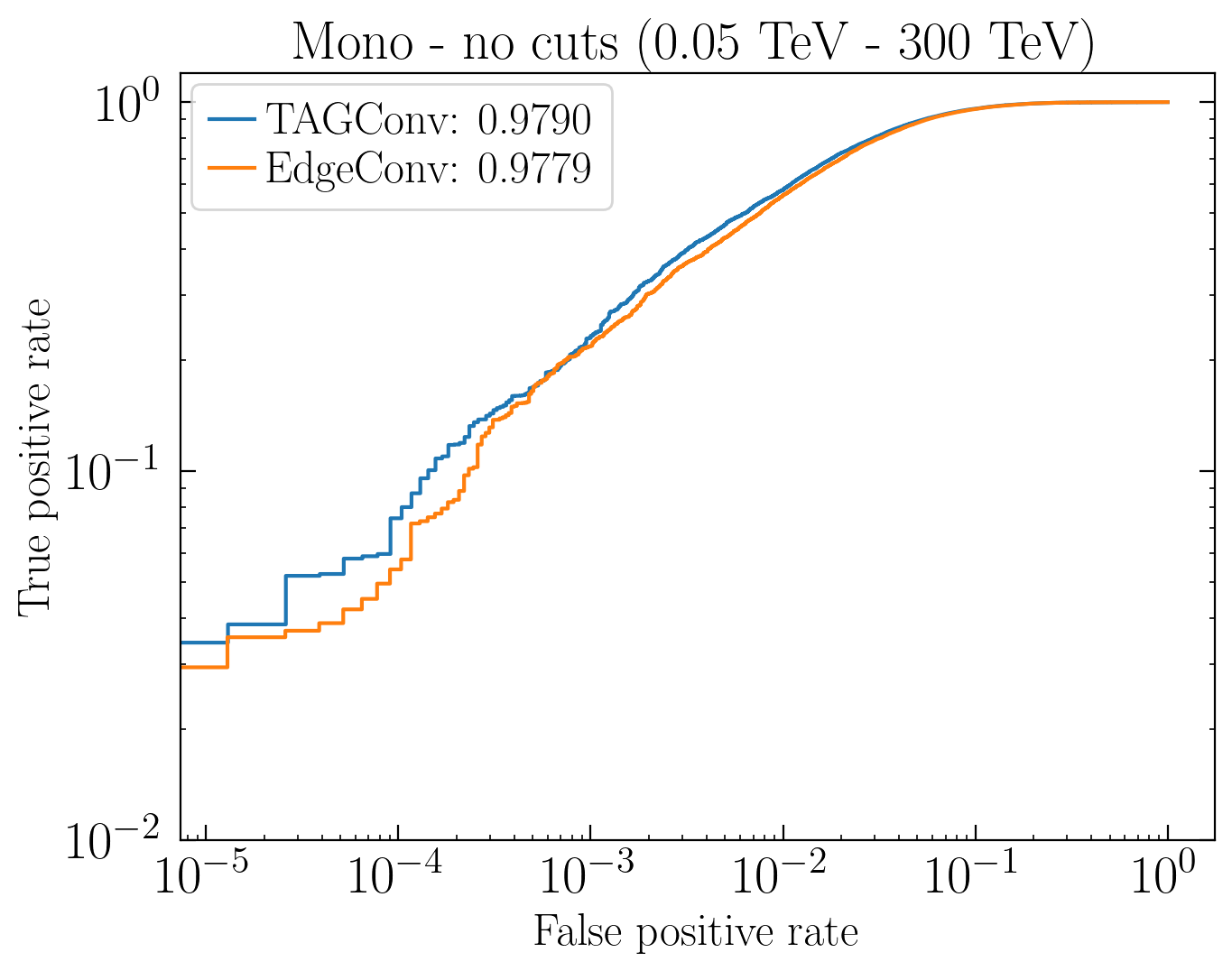}
        \caption{}
        \label{fig:mono_roc}
    \end{subfigure}%
    \begin{subfigure}[b]{0.5\textwidth}
        \centering
        \includegraphics[width=0.99\textwidth]{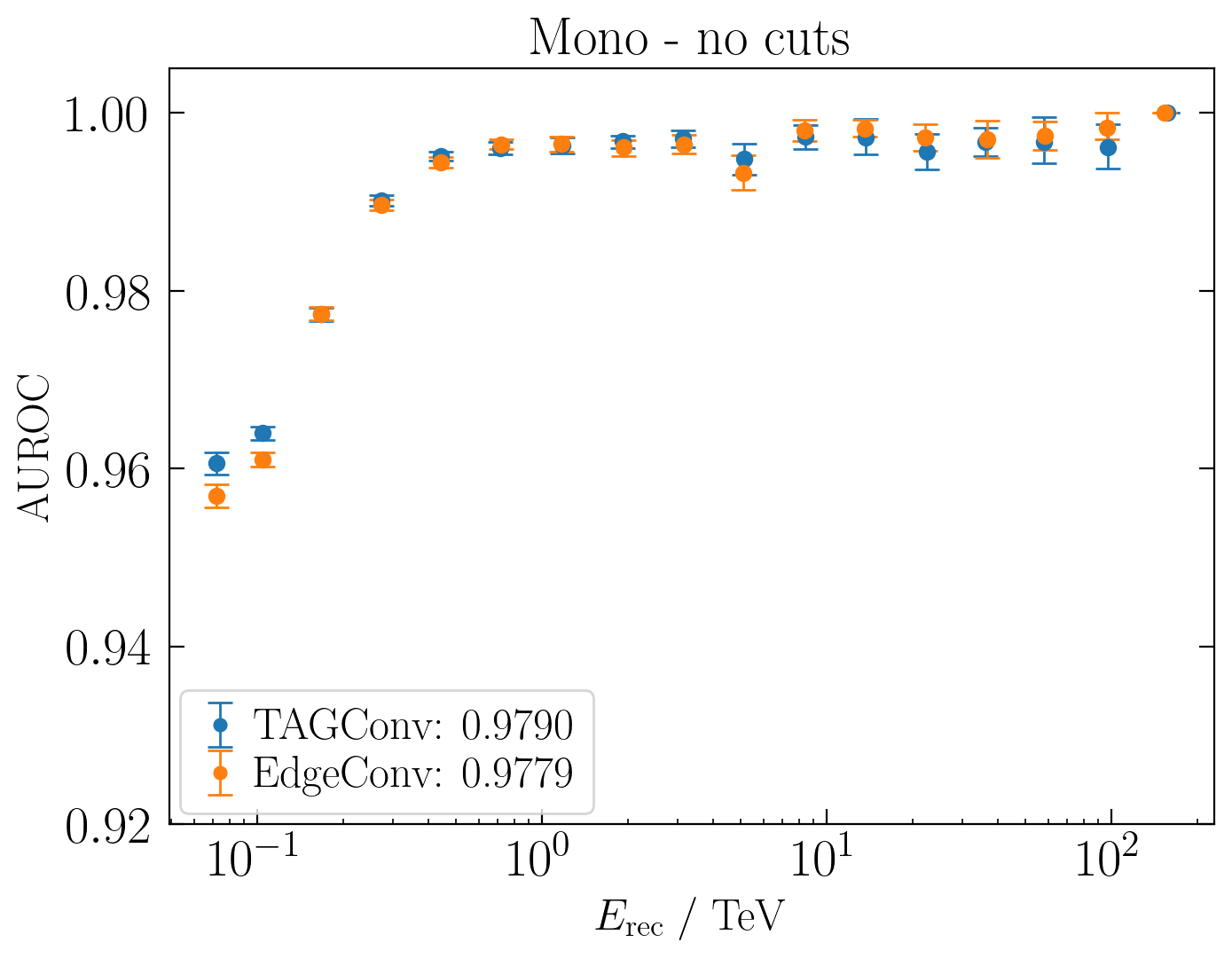}
        \caption{}
        \label{fig:en_dep_mono}
    \end{subfigure}
\caption{Classification performance for mono events without any selection cuts applied. (a) Comparison of ROC curves for the trained graph network classifiers. (b) Energy evolution of the classification performance. The numbers in the legend indicate the AUROC over the whole energy range.}
\label{fig:mono_performance}
\end{figure}

Figure~\ref{fig:mono_performance} shows the performance of our two GCN-based architectures --- TAGConv and EdgeConv --- for the mono configuration, i.e., considering CT5 only. In Figure~\ref{fig:mono_roc}, we examine the general performance over the entire test data set. It can be observed that both architectures give very similar performances of an AUROC 0.9790 for TAGConv and 0.9779 for EdgeConv, respectively. The energy-dependent performance is shown in Figure~\ref{fig:en_dep_mono}. At reconstructed energies larger than 500~GeV, the performance reaches a stable plateau with an AUROC above 0.99 for both architectures.

\begin{figure}[t!]
    \begin{subfigure}[b]{0.5\textwidth}
        \centering
        \includegraphics[width=0.99\textwidth]{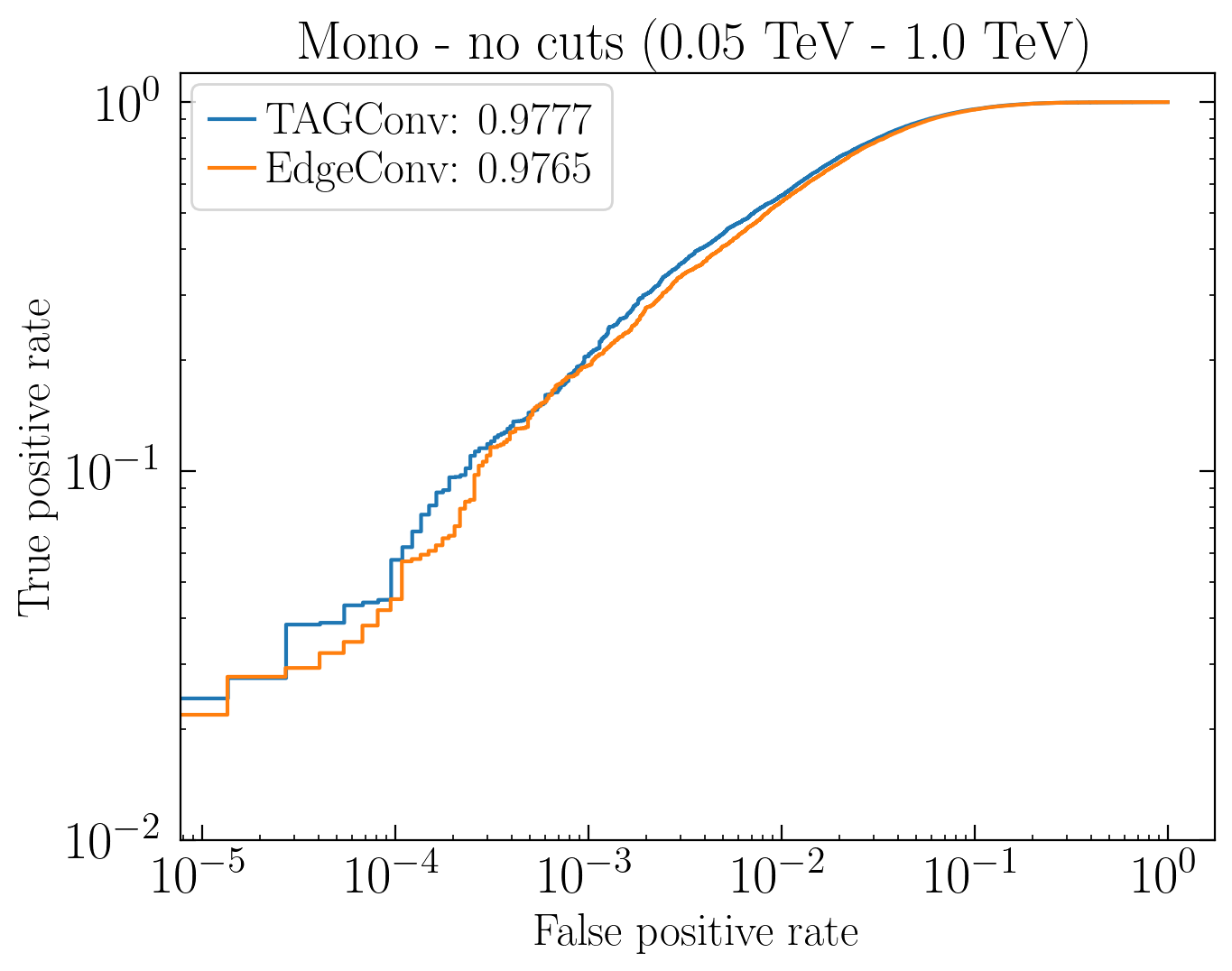}
        \caption{}
        \label{fig:mono_low_energy}
    \end{subfigure}%
    \begin{subfigure}[b]{0.5\textwidth}
        \centering
        \includegraphics[width=0.99\textwidth]{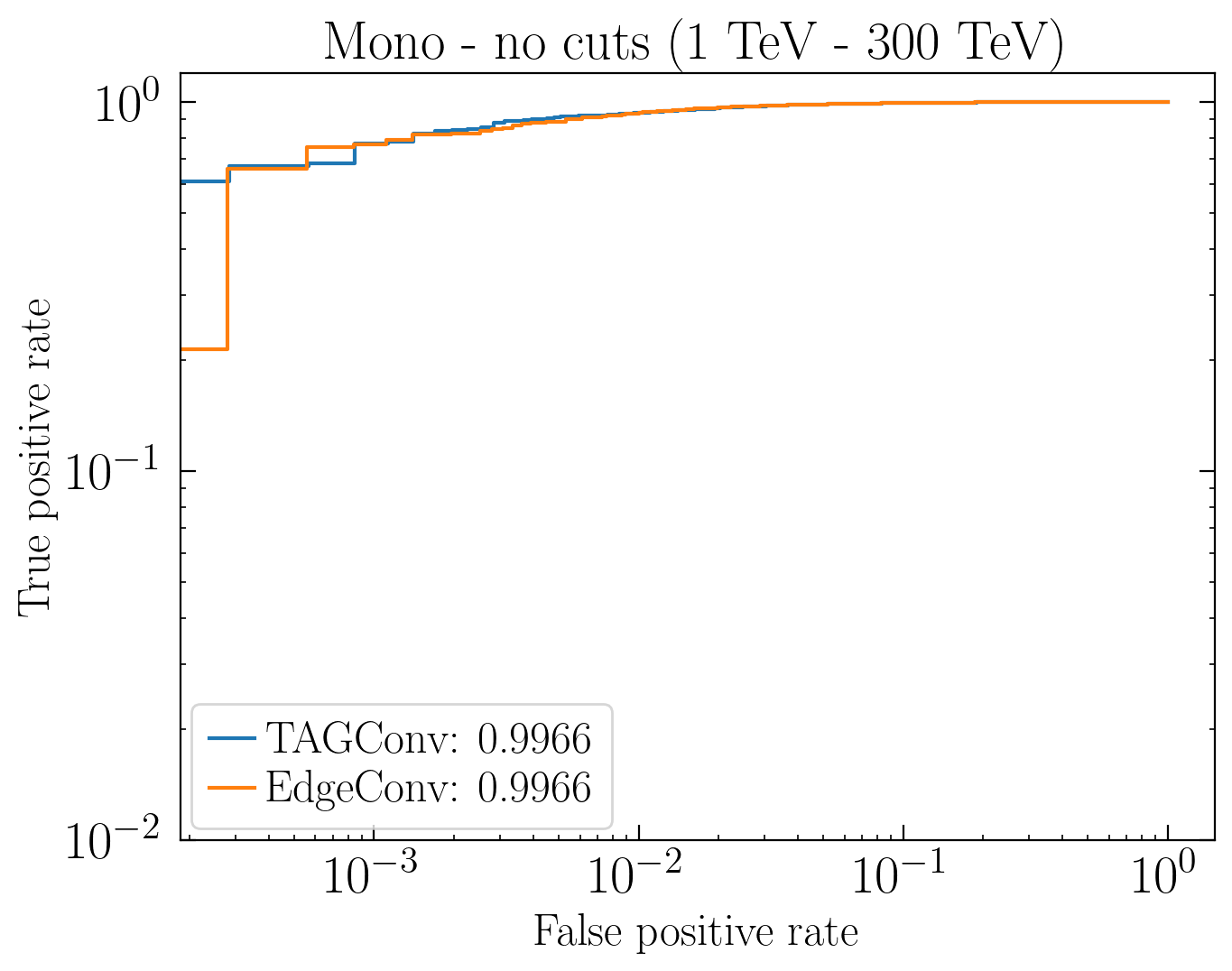}
        \caption{}
        \label{fig:mono_high_energy}
    \end{subfigure}
\caption{Classification performance for mono events for the energy interval (a) 50~GeV to 1~TeV and (b) 1~TeV to 100~TeV.}
\label{fig:mono_performance_energy_bands}
\end{figure}

In Figure~\ref{fig:mono_performance_energy_bands}, we display the performance of our classifiers in two energy bands of 0.05--1~TeV (lower) and 1--300~TeV (higher). At a 50\% of gamma-ray efficiency (0.5 of true positive rate) in the lower energy band, we would misclassify 1 in $\sim100$ ($8\cdot10^{-3}$ false positive rate) background events. Similarly, only 1 in $\sim4000$ background events would be misclassified for the higher energy band. Such good performance of $\gamma$/hadron separation, using a single telescope only, can be attributed to its large collection area, dense pixelization, and the validity of the effectiveness of our deep learning approach. Therefore, the sub-structures in the air-shower light pool can be resolved in more detail, which is especially important for hadron-induced events. Since the mono reconstruction is currently under development, we could not find any available BDT performance of the mono analysis for the H.E.S.S experiment for comparison. 

\subsection{Stereo performance}
\begin{figure}[t!]
    \begin{subfigure}[b]{0.5\textwidth}
        \centering
        \includegraphics[width=0.99\textwidth]{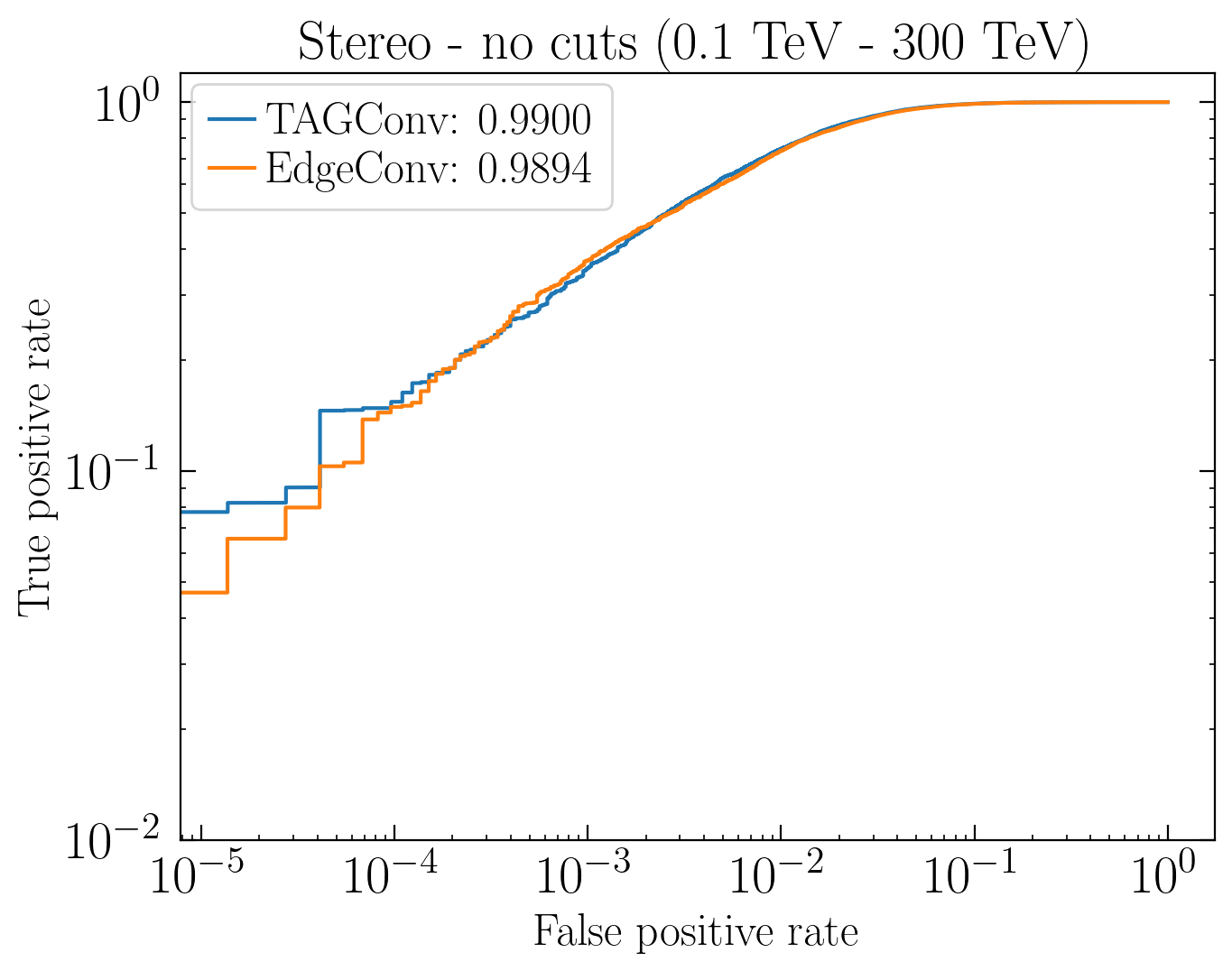}
        \caption{}
        \label{fig:stereo_roc}
    \end{subfigure}%
    \begin{subfigure}[b]{0.5\textwidth}
        \centering
        \includegraphics[width=0.99\textwidth]{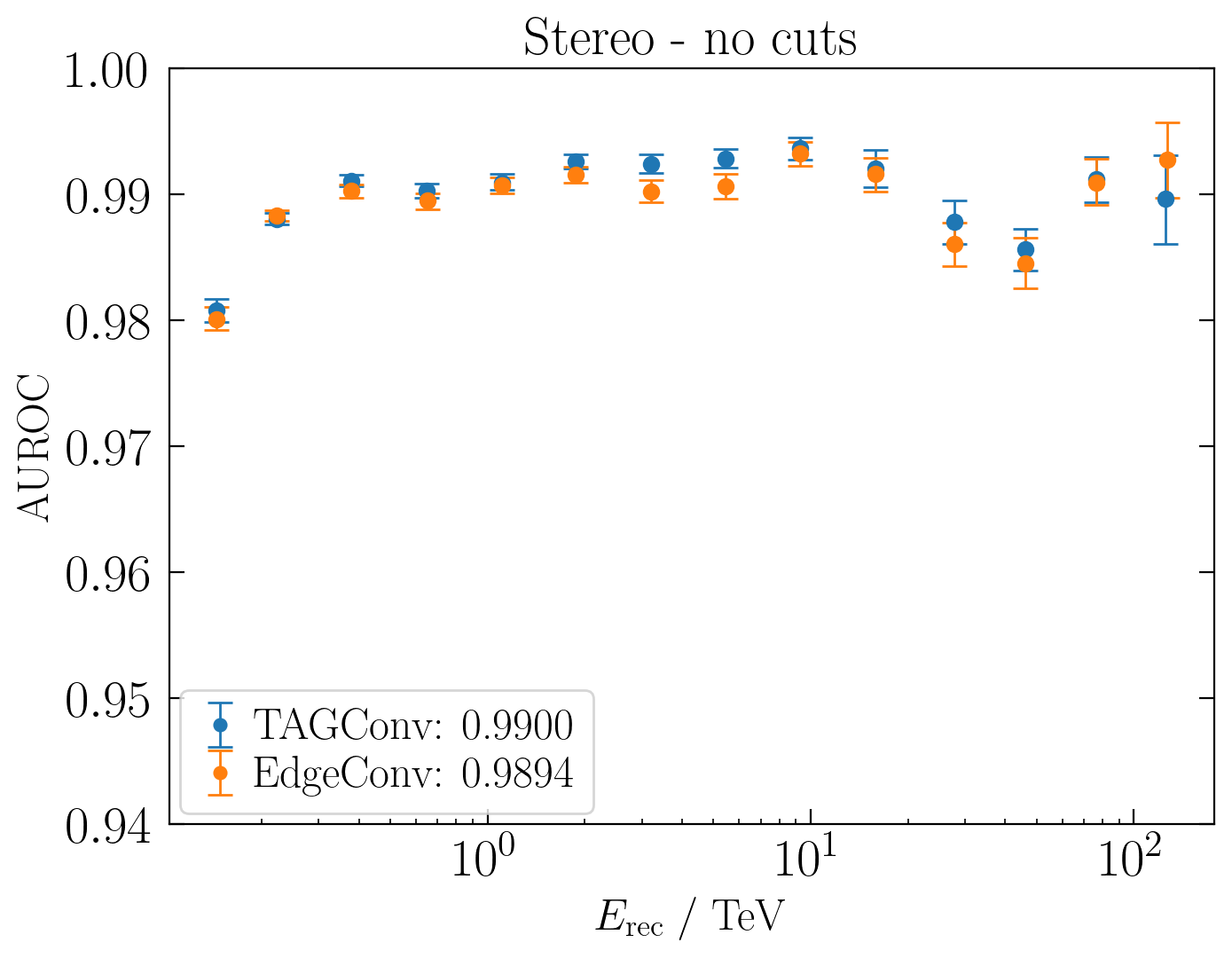}
        \caption{}
        \label{fig:en_dep_stereo}
    \end{subfigure}
\caption{Classification performance for stereo events without preselection cuts applied. (a) ROC curve of the trained graph network classifiers. (b) Energy-dependent classification performance of the graph networks.}
\label{fig:stereo_performance_no_cuts}
\end{figure}

\begin{figure}[t!]
    \begin{subfigure}[b]{0.5\textwidth}
        \centering
        \includegraphics[width=0.99\textwidth]{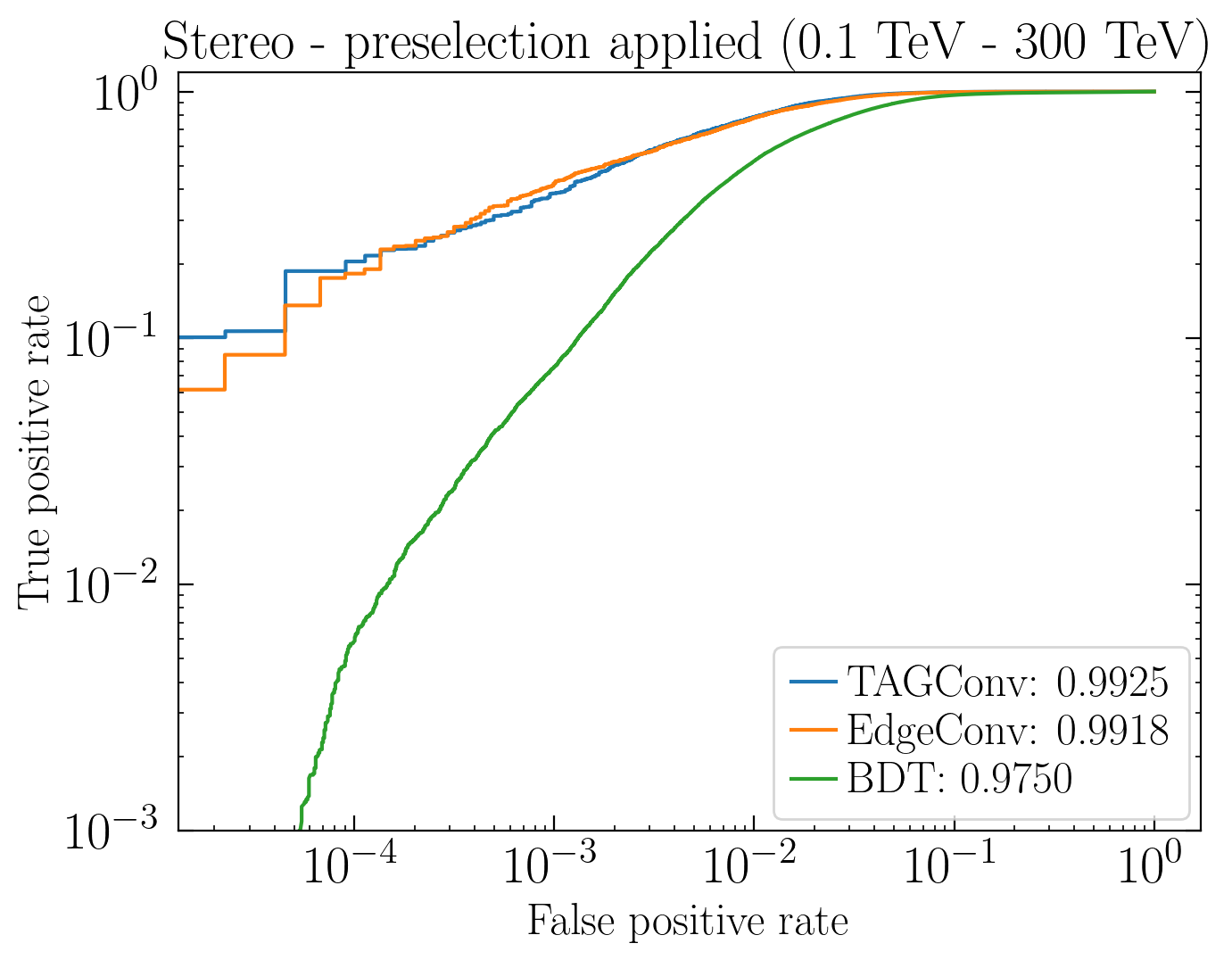}
        \caption{}
        \label{fig:stereo_roc_preselection}
    \end{subfigure}%
    \begin{subfigure}[b]{0.5\textwidth}
        \centering
        \includegraphics[width=0.99\textwidth]{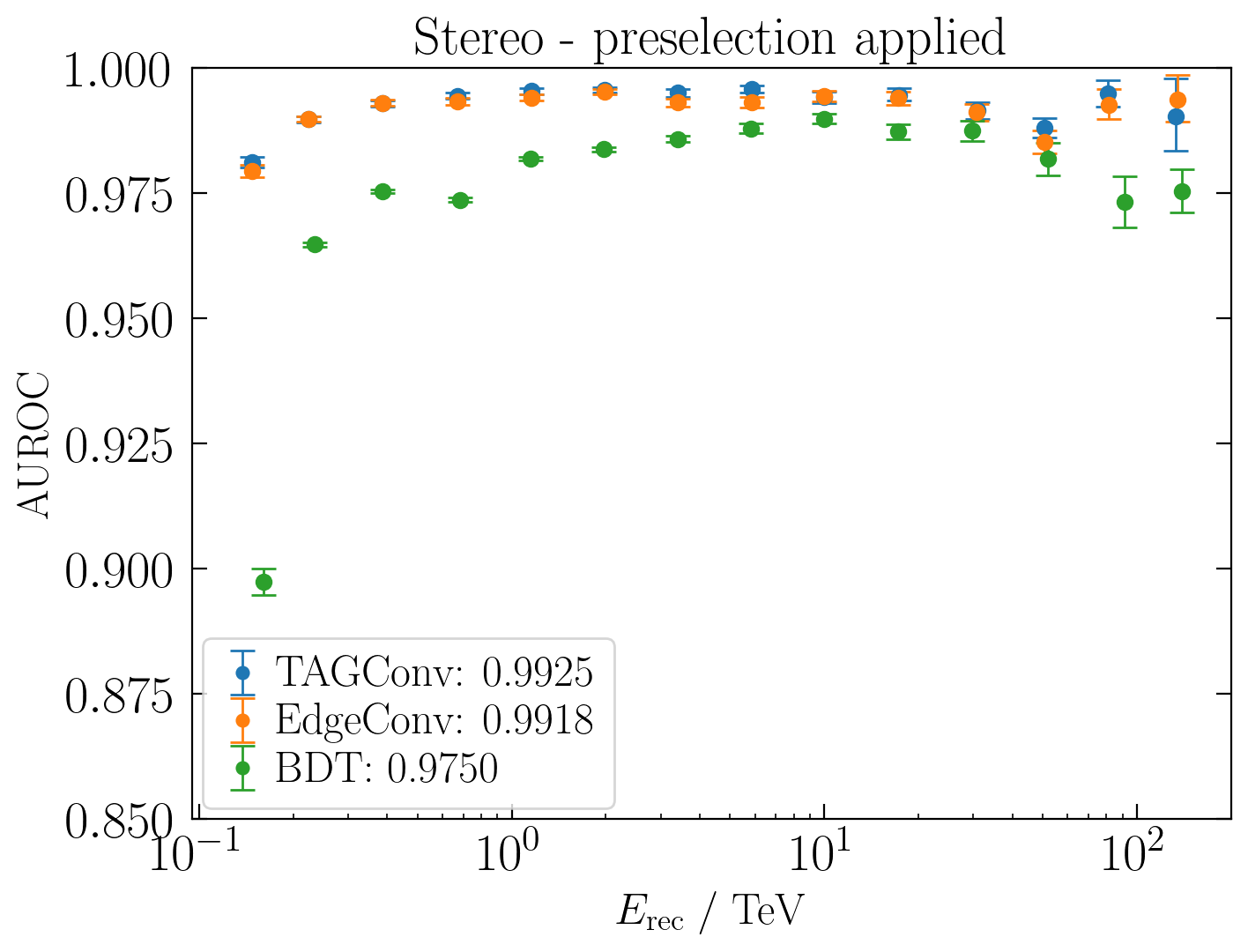}
        \caption{}
        \label{fig:en_dep_stereo_preselection}
    \end{subfigure}
\caption{Classification performance for stereo events with the preselection applied. (a) Comparison of the classification between the two graph networks and the BDT classifier of the standard reconstruction. (b) Energy-dependent classification performance of the classifiers.}
\label{fig:stereo_performance_preselection}
\end{figure}

The $\gamma$/hadron classification performance obtained in this work and its comparison to the BDT-based method for the stereo analysis are shown in Figure~\ref{fig:stereo_performance_no_cuts} and Figure~\ref{fig:stereo_performance_preselection} for the full energy range. In the case of stereo evaluation, it is worth revisiting our cuts applied to the data sets.
We studied the performance for two cases, first: where only the image amplitude cuts are applied (Figure~\ref{fig:stereo_performance_no_cuts}); second: the full preselection cuts are applied, i.e., only events are included that have at least two telescopes from CT1-4 passing the local-distance cut defined in Section~\ref{sec:data} (Figure~\ref{fig:stereo_performance_preselection}). We note that the local distance cut ensures a high-quality reconstruction using the BDT but significantly reduces the statistics by $30\%$.

Figure \ref{fig:stereo_roc} shows the ROC curves obtained for both of our architectures without the preselection cuts. They perform very similarly with an AUROC of 0.9900 and 0.9894 for TAGConv and EdgeConv, respectively. A comparison to the BDT-based classifier could not be shown as in the default H.E.S.S. analysis scheme, the BDTs are only trained and evaluated on events with the preselection applied.
We slightly outperform the CNN-RNN-based study done by Ref.~\cite{Shilon_2019}; they have obtained an AUROC of 0.9875 in this case. Note that AUROC is a highly non-linear metric. Thus, for making a detailed comparison, it would be interesting to compare the proton contamination of the methods at typical working points of background rejection methods in gamma-ray analyses, e.g., at a $\gamma$-efficiency of $0.5$. However, these are unfortunately not available at present.
The AUROC as a function of the reconstructed energy is depicted in Figure~\ref{fig:en_dep_stereo}. It is evident that the performance is very stable in the entire energy range. 

In Figure \ref{fig:stereo_roc_preselection}, we present the ROC curves obtained for preselection cuts applied for both our architectures and the BDT-based method currently used in H.E.S.S. It is to be noted that no re-training was performed on the preselected dataset, but the networks trained on the full data set were used. It is evident that both of our architectures perform very similarly and outperform the BDT-based classifier by a significant margin. The AUROC values obtained for TAGConv and EdgeConv are 0.9925 and 0.9918, respectively, while the BDT-based method only reaches 0.9750.

Note that the improvement only sounds small due to the non-linear behavior of the AUROC. To elaborate on its significance, while comparing the performance of our architectures at $50\%$ gamma-ray efficiency for without and with full preselection cuts, we would misclassify only one background event in $\sim400$ and $\sim700$ events, respectively, in the entire energy range. Similarly, compared to the latter, for the BDT-based method, one in $\sim100$ background events would be misclassified.
This finding signifies a strong performance improvement using the deep learning methods in comparison to the BDT. 
We further marginally outperform (AUROC of 0.9915) the previous CNN study~\cite{Shilon_2019}. In Figure \ref{fig:en_dep_stereo_preselection}, we show the AUROC as a function of reconstructed energy. The difference in the performance of the BDT-based method to ours can be clearly seen below 10~TeV and above 50~TeV.


\begin{figure}[t!]
    \begin{subfigure}[b]{0.5\textwidth}
        \centering
        \includegraphics[width=0.99\textwidth]{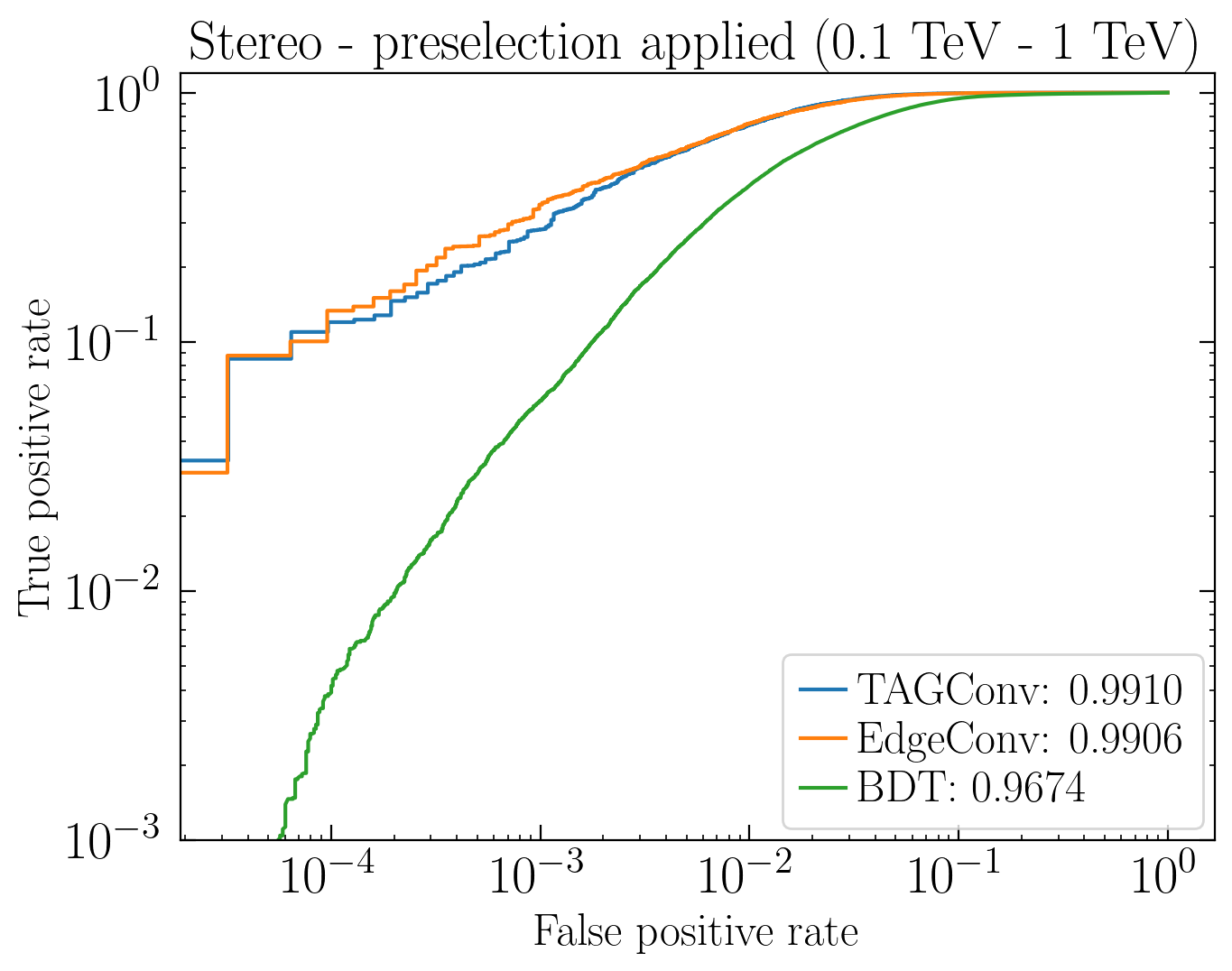}
        \caption{}
        \label{fig:stereo_performance_preselection_energy_bands_low}
    \end{subfigure}%
    \begin{subfigure}[b]{0.5\textwidth}
        \centering
        \includegraphics[width=0.99\textwidth]{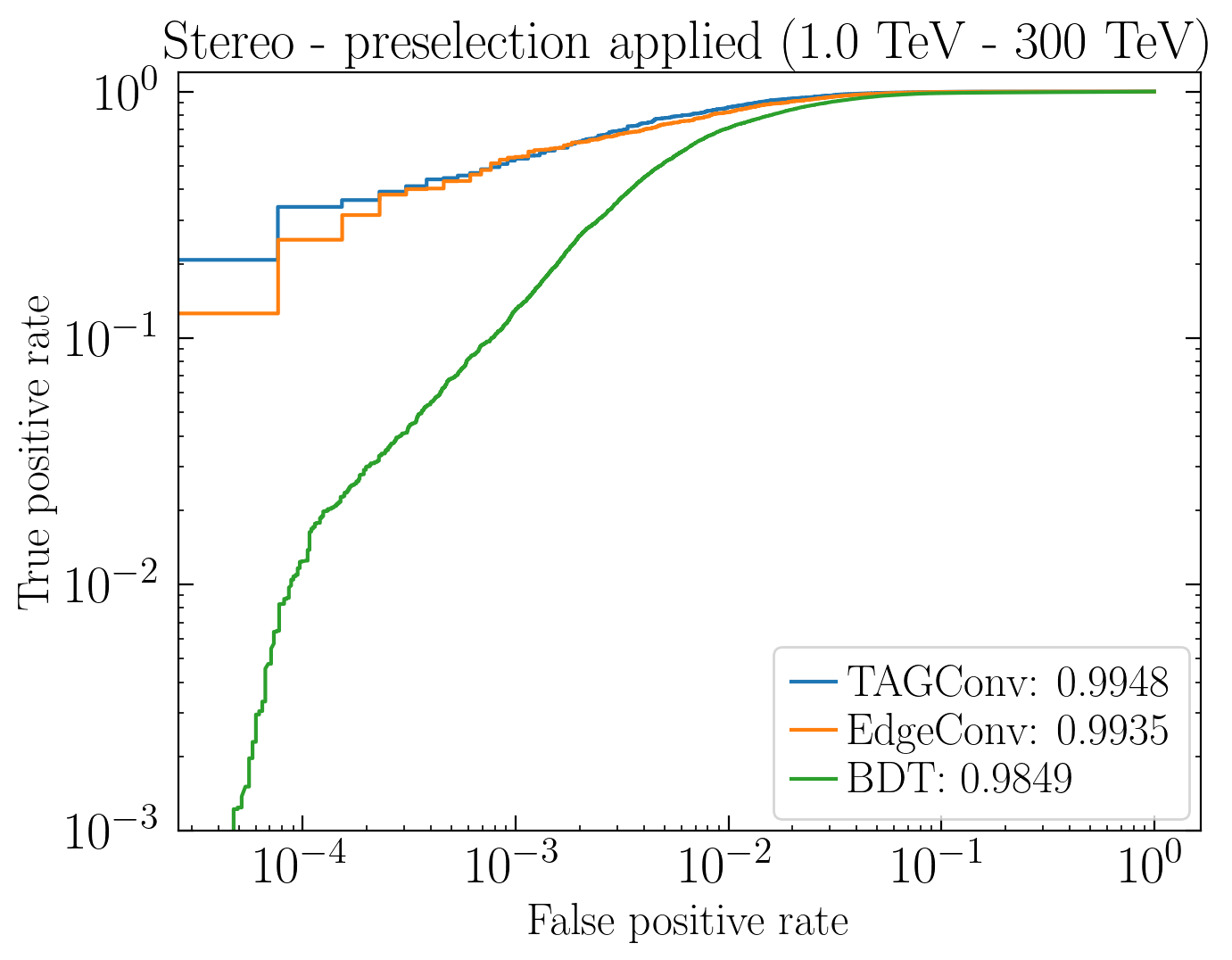}
        \caption{}
        \label{fig:stereo_performance_preselection_energy_bands_high}
    \end{subfigure}
\caption{Classification performance for stereo events with the preselection applied for the energy interval (a) 100 GeV to 1 TeV and (b) 1 TeV to 100 TeV.}
\label{fig:stereo_performance_preselection_energy_bands}
\end{figure}

Finally, in Figure~\ref{fig:stereo_performance_preselection_energy_bands_low} and Figure~\ref{fig:stereo_performance_preselection_energy_bands_high}, we show the ROC curves obtained in two energy bands of 0.1--1~TeV and 1--300~TeV for the cut data set with preselection cuts.
To keep a 50$\%$ $\gamma$-ray efficiency (0.5 of true positive rate), we would misclassify 1 in 500 background events ($2\cdot10^{-3}$ false positive rate) for the energy band of 0.1--1~TeV, which is about an order of magnitude better performance compared to the BDT-based method and a significant improvement. For higher energies above 1~TeV at $50 \%$ $\gamma$-ray efficiency, we would even misclassify only 1 in $\sim1500$ background events, which is again an order of magnitude better than in the case of the BDT-based method.

We note that in~\cite{OHM2009383}, a selection threshold of $>80 \%$ $\gamma$-ray efficiency is used.
However, this threshold is optimized for the BDT-based method using simulations of a point-like source with given properties and optimized to obtain the best detection significance.
This optimization is a trade-off with the Point-Spread-Function (PSF). Performing such optimization is out of the scope of this work.
Therefore, we chose a nominal value of $50 \%$ $\gamma$-ray efficiency. At $80 \%$ $\gamma$-ray efficiency, the improvement achieved in performance would be in a range of three to five instead of an order of magnitude. However, it is worth mentioning that the performance for the BDT-based method is obtained using point-source $\gamma$-ray simulations, while the graph networks are evaluated using diffuse $\gamma$-ray simulations. Hence, a further drop in performance can be expected for the BDT-based classifier as it is comparatively more difficult to classify diffuse $\gamma$-ray simulations than the point source ones from a diffuse proton background.
Due to the aforementioned reasons, it is not straightforward to quantify the improvement achieved more accurately than stated above.

\section{Conclusion}
In this work, we presented a novel algorithm based on graph-convolutional networks for discriminating between photon and hadron-initiated air showers using the images detected by single or multiple Imaging Air Cherenkov Telescopes (IACTs). 
As an example IACT array, we used simulated proton and photon events for the current configuration of H.E.S.S.
By interpreting the cleaned images as graphs, we overcame the inefficient and sparse images in approaches based on convolutional neural networks enabling the application for large IACT arrays featuring of different camera geometries, such as Cherenkov Telescope Array (CTA)~\cite{CTA}.
For two different graph convolutional networks, we have demonstrated that graph networks enable a more natural and efficient classification of IACT images. We studied our approach on two different data sets and found similar results in both methods. For the mono dataset, only the recently-added CT5 telescope was considered. Here, we found a performance enabling a strong separation between protons and photons. In particular, we found a very promising background rejection at low energies (below 1 TeV).\newline
Using the stereo data set, we demonstrated that stereoscopic observations using the CT1 - CT4 telescopes can be analyzed with graph convolutional networks. We found that we can significantly outperform the classical background rejection strategy based on BDTs using deep learning. At a gamma-ray efficiency of $50\%$, the background rejection is improved by roughly one order of magnitude. We further found a promising improvement compared to previous deep-learning-based studies using H.E.S.S. simulation, where we found an improvement in the data with and without preselection cuts.
The finding that the performance of the deep-learning-based algorithms does not strongly rely on the preselection cut is a promising finding. It will enable to increase the statistics in future gamma-ray analyses of up to $50\%$ with an increase in background rejection almost a decade superior to the current generation of classifiers.

This work comprises a novel study of graph networks for IACT images. We expect that using future architecture optimizations, e.g., by including sophisticated attention mechanisms between telescopes to resolve the stereoscopic nature of the images better, improving the image cleaning, and considering pixel-timing will further enhance the performance. Finally, we like to stress that it will be crucial to investigate the difference between data and simulation~\cite{Parsons_2020, parsons2022investigations}, including the night sky background and the instrument response, to exploit the full potential of deep learning algorithms for telescope observations under real-operation conditions.

\acknowledgments
We thank the H.E.S.S. Collaboration for allowing us to use H.E.S.S. simulations for this publication. We further thank Simon Steinmassl and the H.E.S.S. simulation team for running and producing the simulations. The authors would like to thank A.~M.~W. Mitchell and S.~T. Spencer for valuable feedback on the draft of this manuscript.
The authors gratefully acknowledge the scientific support and HPC resources provided by the Erlangen National High Performance Computing Center (NHR@FAU) of the Friedrich-Alexander-Universität Erlangen-Nürnberg (FAU) under the NHR project b129dc. NHR funding is provided by federal and Bavarian state authorities. NHR@FAU hardware is partially funded by the German Research Foundation (DFG) – 440719683.

\bibliographystyle{unsrtnat}
\bibliography{bibliography.bib}{}

\appendix

\newpage
\section{\label{sec:gnn_details}Network architectures}
\subsection{EdgeConv}
\FloatBarrier

\begin{table}[h!]
    \centering
    \begin{centering}
            \begin{tabular}{ r c r r r r }
                \multicolumn{6}{l}{EdgeConv architecture} \\
                \hline\hline
                 & Layer & features & setting & input shape & output shape \\ 
                \hline
                 & EdgeConv & $n_\mathrm{feat}$ & $\aggfn_{j}$: mean & $n_\mathrm{nodes} \times 3$ & $n_\mathrm{nodes} \times n_\mathrm{feat}$ \\ 
                 & SiLU & -- & -- & $n_\mathrm{nodes} \times n_\mathrm{feat}$ & $n_\mathrm{nodes} \times n_\mathrm{feat}$ \\ 
                 & EdgeConv & $n_\mathrm{feat}$ & $\aggfn_{j}$: mean & $n_\mathrm{nodes} \times n_\mathrm{feat}$ & $n_\mathrm{nodes} \times n_\mathrm{feat}$ \\ 
                 & SiLU & -- & -- & $n_\mathrm{nodes} \times n_\mathrm{feat}$ & $n_\mathrm{nodes} \times n_\mathrm{feat}$ \\ 
                 & EdgeConv & $n_\mathrm{feat}$ & $\aggfn_{j}$: mean & $n_\mathrm{nodes} \times n_\mathrm{feat}$ & $n_\mathrm{nodes} \times n_\mathrm{feat}$ \\ 
                 & SiLU & -- & -- & $n_\mathrm{nodes} \times n_\mathrm{feat}$ & $n_\mathrm{nodes} \times n_\mathrm{feat}$ \\ 
                 & EdgeConv & $n_\mathrm{feat}$ & $\aggfn_{j}$: mean & $n_\mathrm{nodes} \times n_\mathrm{feat}$ & $n_\mathrm{nodes} \times n_\mathrm{feat}$ \\ 
                 & SiLU & -- & -- & $n_\mathrm{nodes} \times n_\mathrm{feat}$ & $n_\mathrm{nodes} \times n_\mathrm{feat}$ \\ 
                 & EdgeConv & $n_\mathrm{feat}$ & $\aggfn_{j}$: mean & $n_\mathrm{nodes} \times n_\mathrm{feat}$ & $n_\mathrm{nodes} \times n_\mathrm{feat}$ \\ 
                 & SiLU & -- & -- & $n_\mathrm{nodes} \times n_\mathrm{feat}$ & $n_\mathrm{nodes} \times n_\mathrm{feat}$ \\ 
                 & EdgeConv & $n_\mathrm{feat}$ & $\aggfn_{j}$: mean & $n_\mathrm{nodes} \times n_\mathrm{feat}$ & $n_\mathrm{nodes} \times n_\mathrm{feat}$ \\ 
                 & SiLU & -- & -- & $n_\mathrm{nodes} \times n_\mathrm{feat}$ & $n_\mathrm{nodes} \times n_\mathrm{feat}$ \\ 
                 & Concat & -- & layer output & $ 6 \cdot (n_\mathrm{nodes} \times n_\mathrm{feat}$) & $n_\mathrm{nodes} \times 6\cdot n_\mathrm{feat}$ \\
                 & Concat & -- & telescopes & $n_\mathrm{tel} \cdot (n_\mathrm{nodes} \times 6\cdot n_\mathrm{feat})$ & $n_\mathrm{nodes} \times 6\cdot n_\mathrm{feat}\cdot n_\mathrm{tel}$ \\
                 & MaxPooling & -- & -- & $n_\mathrm{nodes} \times 6\cdot n_\mathrm{feat} \cdot n_\mathrm{tel}$ & $6\cdot n_\mathrm{feat} \cdot n_\mathrm{tel}$ \\
                 & Batchnorm & -- & -- & $n_\mathrm{nodes} \times 6\cdot n_\mathrm{feat} \cdot n_\mathrm{tel}$ & $6\cdot n_\mathrm{feat} \cdot n_\mathrm{tel}$ \\
                 & Dropout & -- & $p=0.5$ & $6\cdot n_\mathrm{feat} \cdot n_\mathrm{tel}$ & $6\cdot n_\mathrm{feat} \cdot n_\mathrm{tel}$ \\
                 & Linear & $n_{\mathrm{feat}}$ & -- & $6\cdot n_\mathrm{feat} \cdot n_\mathrm{tel}$ & $n_{\mathrm{feat}}$  \\
                 \multirow{2}{*}{$2 \times \Big{\{}$} & ResNet & $n_{\mathrm{feat}}$ & -- & $n_{\mathrm{feat}}$ & $n_{\mathrm{feat}}$ \\
                 & Dropout & -- & $p=0.5$ & $n_{\mathrm{feat}}$ & $n_{\mathrm{feat}}$ \\
                 & Linear & $2$ & -- & $n_{\mathrm{feat}}$ & $2$ \\
                 & Softmax & -- & -- & 2 & $2$ \\

                \hline
            \end{tabular}
            \caption{EdgeConv architecture for the analysis of mono and stereo IACT images. $n_\mathrm{nodes}$ denotes the number of nodes in the input graph, $n_\mathrm{tel}$ the number of telescopes (one for mono, four for stereo), and $n_{\mathrm{feat}}$ the number of features, which is $256$ for the mono and $96$ for the stereo dataset.}
            \label{tab:edge_conv_arch}
    \end{centering}
\end{table}

\FloatBarrier

\begin{table}[bh!]
    \begin{centering}
        \begin{subtable}[h]{0.495\textwidth}
            \begin{tabular}{ c r r }
                \multicolumn{3}{l}{Kernel network $h_\mathbf{\Theta}$} \\
                \hline\hline
                layer & features & settings \\
                \hline
                Linear & $n_\mathrm{feat}$ & no bias \\
                Batchnorm & -- & momentum=0.9  \\
                SiLU & --  & --  \\
                Linear & $n_\mathrm{feat}$ & no bias \\
                Batchnorm & -- & momentum=0.9  \\
                SiLU & -- & -- \\
                Linear & $n_\mathrm{feat}$ & no bias \\
                Batchnorm & -- & momentum=0.9  \\
                SiLU & -- & -- \\
                \hline
            \end{tabular}
            \caption{Architecture of the kernel function $h_\mathbf{\Theta}$, where $n_\mathrm{feat}$ is to be chosen by the user. For each layer, a different kernel function was used.}
            \label{tab:edge_conv_ffn}
        \end{subtable}
        \hfill
        \begin{subtable}[h]{0.495\textwidth}
            \begin{centering}
                \begin{tabular}{ c r r }
                    \multicolumn{3}{l}{Residual Module "ResNet"}\\
                    \hline\hline
                    layer & features & settings \\
                    \hline
                    Linear & $n_\mathrm{feat}$ & no bias \\
                    SiLU & --  & --  \\
                    Batchnorm & -- & -- \\
                    Linear & $n_\mathrm{feat}$ & no bias \\
                    SiLU & --  & --  \\
                    Batchnorm & -- & -- \\
                    Add & -- & -- \\
                    SiLU & -- & -- \\
                    \hline
                \end{tabular}
                \caption{Details of our used ResNet modules.}
                \label{tab:resnet_arch}
           \end{centering}
        \end{subtable}
    \end{centering}
    \label{tab:edge_conv_kernel}
\end{table}

\FloatBarrier

\newpage

\subsection{TAGConv}
\FloatBarrier

\begin{table}[h!]
    \centering
    \begin{centering}
            \begin{tabular}{ r c r r r r }
                \multicolumn{6}{l}{TAGConv architecture} \\
                \hline\hline
                 & Layer & features & setting & input shape & output shape \\ 
                \hline
                 & TAGConv & $n_\mathrm{feat}$ & agg: sum & $n_\mathrm{nodes} \cdot 3$ & $n_\mathrm{nodes} \cdot n_\mathrm{feat}$ \\ 
                 & ReLU & -- & -- & $n_\mathrm{nodes} \cdot n_\mathrm{feat}$ & $n_\mathrm{nodes} \cdot n_\mathrm{feat}$ \\ 
                 & TAGConv & $n_\mathrm{feat}$ & agg: sum & $n_\mathrm{nodes} \cdot n_\mathrm{feat}$ & $n_\mathrm{nodes} \cdot n_\mathrm{feat}$ \\ 
                 & ReLU & -- & -- & $n_\mathrm{nodes} \cdot n_\mathrm{feat}$ & $n_\mathrm{nodes} \cdot n_\mathrm{feat}$ \\ 
                 & TAGConv & $n_\mathrm{feat}$ & agg: sum & $n_\mathrm{nodes} \cdot n_\mathrm{feat}$ & $n_\mathrm{nodes} \cdot n_\mathrm{feat}$ \\ 
                 & ReLU & -- & -- & $n_\mathrm{nodes} \cdot n_\mathrm{feat}$ & $n_\mathrm{nodes} \cdot n_\mathrm{feat}$ \\ 
                 & TAGConv & $n_\mathrm{feat}$ & agg: sum & $n_\mathrm{nodes} \cdot n_\mathrm{feat}$ & $n_\mathrm{nodes} \cdot n_\mathrm{feat}$ \\ 
                 & ReLU & -- & -- & $n_\mathrm{nodes} \cdot n_\mathrm{feat}$ & $n_\mathrm{nodes} \cdot n_\mathrm{feat}$ \\ 
                 & TAGConv & $n_\mathrm{feat}$ & agg: sum & $n_\mathrm{nodes} \cdot n_\mathrm{feat}$ & $n_\mathrm{nodes} \cdot n_\mathrm{feat}$ \\ 
                 & ReLU & -- & -- & $n_\mathrm{nodes} \cdot n_\mathrm{feat}$ & $n_\mathrm{nodes} \cdot n_\mathrm{feat}$ \\ 
                 & TAGConv & $n_\mathrm{feat}$ & agg: sum & $n_\mathrm{nodes} \cdot n_\mathrm{feat}$ & $n_\mathrm{nodes} \cdot n_\mathrm{feat}$ \\ 
                 & ReLU & -- & -- & $n_\mathrm{nodes} \cdot n_\mathrm{feat}$ & $n_\mathrm{nodes} \cdot n_\mathrm{feat}$ \\ 
                 & Concat & -- & layer output & $ 6 \cdot (n_\mathrm{nodes} \cdot n_\mathrm{feat})$ & $n_\mathrm{nodes} \cdot 6 \cdot n_\mathrm{feat}$ \\
                 & Concat & -- & telescopes & $n_\mathrm{tel} \cdot (n_\mathrm{nodes} \cdot 6 \cdot n_\mathrm{feat})$ & $n_\mathrm{nodes} \cdot 6 \cdot n_\mathrm{feat} \cdot n_\mathrm{tel}$ \\
                 & MaxPooling & -- & -- & $n_\mathrm{nodes} \cdot 6 \cdot n_\mathrm{feat} \cdot n_\mathrm{tel}$ & $6 \cdot n_\mathrm{feat} \cdot n_\mathrm{tel}$ \\
                 & Batchnorm & -- & -- & $n_\mathrm{nodes} \cdot 6\cdot n_\mathrm{feat} \cdot n_\mathrm{tel}$ & $6\cdot n_\mathrm{feat} \cdot n_\mathrm{tel}$ \\
                 & Dropout & -- & $p=0.5$ & $6 \cdot n_\mathrm{feat} \cdot n_\mathrm{tel}$ & $6 \cdot n_\mathrm{feat} \cdot n_\mathrm{tel}$ \\
                 & Linear & $n_{\mathrm{feat}}$ & -- & $6 \cdot n_\mathrm{feat} \cdot n_\mathrm{tel}$ & $n_{\mathrm{feat}}$  \\
                 \multirow{2}{*}{$2\cdot \Big{\{}$} & ResNet & $n_{\mathrm{feat}}$ & -- & $n_{\mathrm{feat}}$ & $n_{\mathrm{feat}}$ \\
                 & Dropout & -- & $p=0.5$ & $n_{\mathrm{feat}}$ & $n_{\mathrm{feat}}$ \\
                 & Linear & $2$ & -- & $n_{\mathrm{feat}}$ & $2$ \\
                 & Softmax & -- & -- & 2 & $2$ \\

                \hline
            \end{tabular}
            \caption{TAGConv architecture used in this work to analyze H.E.S.S. mono and stereo analysis configuration. $n_{\mathrm{feat}}$ denotes the number of features which is 150 for stereo and 130 for mono. "agg" means aggregation.  $n_\mathrm{nodes}$ corresponds to the number of nodes in the input graph, and $n_\mathrm{tel}$ is the number of telescopes (one for mono, four for stereo). The "ResNet" is the same as shown in Table~\ref{tab:resnet_arch}.}
            \label{tab:tagconv_arch}
    \end{centering}
\end{table}
\FloatBarrier

\subsection{Training parameters}
\paragraph{Mono}
For training the EdgeConv model on the mono dataset ($n_\mathrm{tel} = 1$), a batch size of 128 samples was used with an initial learning rate of $1 \cdot 10^{-3}$. Further, we set $n_\mathrm{feat}=256$. For the TAGConv model, we use a batch size of 96 samples with an initial learning rate of $5 \cdot 10^{-3}$. The number of features for the TAGConv model was 130. Additionally, for both models learning rate was reduced, by multiplying with $\delta = 0.33$, when the validation loss did not decrease after five epochs. The training was stopped after the validation loss did not decrease after ten epochs.

\paragraph{Stereo}
For training the EdgeConv model on the stereo dataset ($n_\mathrm{tel} = 4$), a batch size of 96 samples was used with an initial learning rate of $10^{-3}$. Further, we set $n_\mathrm{feat}=96$. For the TAGConv model, we use a batch size of 96 samples with an initial learning rate of $5 \cdot 10^{-3}$. The number of features for the TAGConv model was 150. Additionally, for both models, the learning rate was reduced by multiplying with $\delta = 0.3$ for EdgeConv, and $\delta = 0.33$ for TAGConv, when the validation loss did not decrease after five epochs. The training was stopped after the validation loss did not decrease after ten epochs.

\end{document}